\documentclass[11pt]{article}
\usepackage{amssymb,epsf,graphicx}
\usepackage{amsmath}
\allowdisplaybreaks[1]

\newcommand{\be}{\begin{equation}}
\newcommand{\ee}{\end{equation}}
\newcommand{\ben}{\begin{eqnarray}\displaystyle}
\newcommand{\een}{\end{eqnarray}}
\newcommand{\sectiono}[1]{\section{#1}\setcounter{equation}{0}}

\newcommand\crbig{\\\noalign{\vspace {1.5mm}}}

\def\Im{\mathop{\rm Im}}
\def\Re{\mathop{\rm Re}}

\newcommand{\vac}{|0\rangle}

\begin{document}
{}~ \hfill\vbox{\hbox{hep-th/0609209}\hbox{SISSA 57/2006/EP}}\break
\vskip 2.1cm

\centerline{\large \bf Closed Bosonic String Field Theory at Quintic Order:}
\centerline{\large \bf Five-Tachyon Contact Term and Dilaton Theorem}
\vspace*{8.0ex}

\centerline{\large \rm Nicolas Moeller}

\vspace*{8.0ex}

\centerline{\large \it International School for Advanced Studies (SISSA)}
\centerline{\large \it via Beirut 2-4,}
\centerline{\large \it 34014 Trieste, Italy} \vspace*{2.0ex}
\centerline{E-mail: {\tt moeller@sissa.it}}

\vspace*{6.0ex}

\centerline{\bf Abstract}
\bigskip

We solve the geometry of the closed string field theory five-point
vertex. Our solution is calculated in terms of quadratic Strebel
differentials which are found numerically all over the relevant
subspace of the moduli space of spheres with five punctures. Part of
the boundary of the reduced moduli space is described in terms of an
algebraic curve, while the remaining part has to be evaluated
numerically. We use this data to compute the contact term of five
tachyons and estimate its uncertainty to be of about $0.1\%$. To put
to a test the theory and the computations done, we calculate the
contact term of five dilatons. In agreement with the dilaton theorem,
it is found to cancel the term obtained from the tree level Feynman
diagrams built with three- and four-vertices. This cancellation,
achieved with a precision of about $0.1 \%$, is within the estimated
margin error on the contact term and is therefore a very good evidence
that our computations are reliable. The techniques and numerical
algorithm developed in this paper make it possible to compute the
contact amplitude of any five off-shell closed bosonic string states.

\vfill \eject

\baselineskip=16pt

\tableofcontents


\sectiono{Introduction}
\label{setup_s5}

The object of this paper is the explicit computation of the quintic
term of the action of closed bosonic string field theory (CSFT). This
action, formally constructed in \cite{Zwie-BV,CSFT,Saad-Zwie,area}, is
nonpolynomial. This is in contrast with Witten's string field theory
\cite{Witten:1985cc} which is cubic, i.e. Feynman diagrams constructed
with three-vertices are enough to cover (exactly once) the whole
moduli space of Riemann surfaces with $N$ punctures on the
boundary. In CSFT it is not possible to do so, The Feynman diagrams
with closed three-vertices do not suffice to construct all spheres
with four punctures, one has to introduce, in the action, a contact
term of order four in the string field to account for the remaining
four-punctured spheres. But this is not yet enough, one has to
introduce also a five-vertex because the Feynman diagrams with three-
and four-vertices do not cover the whole moduli space of spheres with
five punctures. And so on, one must put in the action, contact terms
of all orders. In this paper we will discuss only the classical
action, but if we were to consider diagrams with loops, we would face
a similar problem; namely the vertices of the classical action are not
enough to generate all Riemann surfaces of genus one; and so on, one
must introduce terms at all genii.

Being able to explicitly compute the whole action would be immensely
useful in understanding nonperturbative physics of closed strings. The
hot problem that we have in mind is, in particular, to understand if
closed bosonic string theory has a stable vacuum, and what would the
theory look like in this vacuum. The approach that we are taking, is
by order truncation of the CSFT action, i.e. we truncate it to a
polynomial of a given order in the string field. The simplest
nontrivial truncation is done by keeping the quadratic and cubic terms
only. In \cite{Kostelecky:1990mi}, Kosteleck\'y and Samuel did
precisely that; they then truncated the string field itself by keeping
only the tachyon and all the massless fields. In this approximation
they found a locally stable vacuum in which the tachyon had a positive
expectation value.

The next order of approximation is to keep the quartic term as
well. The computation of this term is already seriously
complicated. The reason lies in the fact that, in order to compute a
contact amplitude with $N$ external states, one has to integrate a
certain correlator over a region of the moduli space of spheres with
$N$ punctures (we call this region the {\em reduced} moduli space, it
corresponds to all spheres that cannot be constructed with Feynman
diagrams), which has real dimensionality $2(N-3)$. The expression of
the correlator at a point in the reduced moduli space, depends on the
geometry of the vertex at this point, which in turn is given by the
solution of a minimal area problem \cite{area}. For the cubic vertex
$N=3$, there is no moduli space to integrate over and it is thus easy
to calculate. But for the quartic vertex, there is a two-dimensional
reduced moduli space. This vertex was solved numerically in
\cite{quartic}. The solution given there consisted of the boundary of
the reduced moduli space in the complex plane, and everywhere in this
region the geometry of the vertex was expressed with a quadratic
differential (see \cite{Strebel, Belo-Zwie} for details on quadratic
differentials) given in terms of a complex parameter $a(\xi,
\bar{\xi})$ depending on the coordinates on the reduced moduli space.
The solution was explicitly given by a reasonably short fit that can
be copied from the paper and used to compute amplitudes with an
accuracy of about $0.1\%$.

The results of \cite{quartic} were checked by Yang and Zwiebach in
\cite{Yang:2005iu, Yang:2005ep}. For this, they verified that the
quartic term in the effective potential of some marginal fields is
seen to vanish as one increases the truncation level of the string
field. A similar analysis was made for the effective potential of the
dilaton. In particular, they found that the contact quartic term
cancels the terms from cubic vertices with a precision of about
$0.2\%$.

In \cite{Yang:2005rx,Yang:2005rw}, Yang and Zwiebach went on to
address the question raised before, whether closed bosonic string
theory has a stable vacuum. They started by realizing that the tachyon
condensate {\em must} drive the zero-momentum ghost dilaton. Indeed
this state with a peculiar ghost structure, given by
\be
|D\rangle = \left(c_1c_{-1}-\bar{c}_1 \bar{c}_{-1}\right) |0\rangle
\label{dilaton-field}
\ee 
has to be included in the condensate as soon as one considers the
quartic vertex.\footnote{Several years before the quartic vertex was
solved in \cite{quartic}, Belopolsky \cite{Belo} managed to calculate
the tachyon effective potential to order four and he found that it had
no minimum. This result, however, didn't take into account the
dilaton. In fact, it is understood now that the tachyon effective
potential doesn't make much sense because one cannot integrate out the
massless dilaton.}  This stems from the fact that the antighost
insertion in the correlator, can make the ghost numbers work so that
amplitudes with a single dilaton, for example three tachyons and a
dilaton, are nonzero. They truncated the string field to level four,
including the tachyon (level zero), the dilaton (level two) and four
massive scalar fields at level four. And they found that the closed
string field theory potential has a local minimum where both the
tachyon and dilaton take positive expectation values. They also
noticed that the depth of this minimum tended to decrease as the level
increases, and they made the proposition that this vacuum should have
a vanishing action density. They supported this claim by looking at
the low-energy effective action of the tachyon, dilaton and metric.
They found that if this action has a stable vacuum, its depth must be
zero. Although the numerics to level four seemed to confirm this
claim, a recent computation to level ten \cite{Moe-Yang} shows that,
at quartic order, the value of the potential at the stable vacuum is
actually negative and non-zero. The question now is whether higher
order terms in the CSFT action can make the shallowness of the
potential go to zero, or if it stays finite.

It is clear at this point that we need the quintic term of the CSFT
action. In the present paper we solve numerically the geometry of the
quintic vertex. This is again done with quadratic differentials. We
present in details the algorithm to solve the Strebel condition and we
spend quite some time describing the reduced moduli space of spheres
with five punctures. It will turn out that we can split it into 120
regions, and need to describe only one of them, that we call ${\cal
A}_5$. We undertake the description of the boundary of ${\cal A}_5$,
and to our pleasant surprise we find that its projection on one of its
two complex coordinates (corresponding to the two unfixed punctures in
the uniformizer coordinate) can be described algebraically in terms of
an algebraic curve, that will be found from the solution of the class
of quadratic differentials with two double zeros. The rest of the
boundary will be solved numerically. After this is done, we can
integrate correlators over the reduced moduli space. The simplest one
is the term with five tachyons. We describe in details how we do this
integration and how we estimate the uncertainty in the result.

In order to gain confidence in this result, and \`a fortiori in the
machinery developed and the numerical results produced, we must check
our algorithm in some way. For this, we compute the effective
potential of the dilaton to order five. As the dilaton theorem claims,
this should be identically zero. The term of order five is composed of
two terms, namely the contact term that we calculate with our
algorithm, and the term from Feynman diagrams with vertices of lower
order, that we calculate with the techniques and results of
\cite{Yang:2005ep, Moe-Yang}. The cancellation is achieved with an
accuracy of about $0.1\%$, falling well within the $0.5\%$ estimated
error on the five-dilaton contact term. We therefore claim that our
algorithm and result for the five-tachyon term, are reliable. The
computation of terms of higher levels, necessary in order to pursue
the study of the nonperturbative vacuum of \cite{Yang:2005rx,
Moe-Yang}, is now possible and will be done in a future publication
\cite{Moe-prog}.

We think it is of interest to give some orders of magnitude related to
the algorithm developed here. Its implementation on a computer is done
in the C++ language, and the code is more than 10,000 lines long. The
complete and accurate computation of the moduli space and the
quadratic differentials inside it, takes several days to compute on a
desktop computer and generates about one GByte of data. Once this is
done, the accurate computation of an integral takes several
hours. Unfortunately, the size of the data makes it impossible for now
to express the numerical solution in terms of a reasonably short fit,
as we did for the quartic vertex \cite{quartic}.

\paragraph{}
The paper is structured as follows: We end this section by a short
summary on the CSFT action. In Section \ref{regular_s5}, we construct
the quadratic differentials pertinent to the five-point vertex, and we
study some of their limiting cases in Section \ref{Limits_s5}. We then
describe in Section \ref{solving_s5} how to solve the Strebel
condition numerically. The reduced moduli space and the way to compute
it are described in Section \ref{ms_s5}. We can then integrate to
obtain the first result, namely the five-tachyon contact term, this is
done in Section \ref{s_t5}. The reliability check with the computation
of the dilaton effective potentials is done in Section \ref{s_d5}. We
end with some discussions on the present results and prospects in
Section \ref{s_conclusions}.

\paragraph{}
We now summarize the CSFT action and fix our notation. 
With $\alpha'=2$, the closed string field theory action is \cite{Zwie-BV, CSFT}
\be
S = -\frac{1}{\kappa^2} \left( \frac{1}{2} \langle \Psi | c_0^- Q_B |\Psi \rangle + 
\frac{1}{3!} \left\{\Psi, \Psi, \Psi \right\} + 
\frac{1}{4!} \left\{\Psi, \Psi, \Psi, \Psi \right\} + 
\frac{1}{5!} \left\{\Psi, \Psi, \Psi, \Psi, \Psi \right\} + \ldots \right) \,,
\ee
where $c_0^- = \frac{1}{2} (c_0-\bar{c}_c)$, $Q_B$ is the BRST charge and 
$\{\ldots\}$, are the multilinear string functions. In this paper we won't need 
to know much about the string field $\Psi$, except that 
when considering tachyon condensation we keep only scalars with zero-momentum. 
Namely
\be
|\Psi\rangle = t \, |T\rangle + d \, |D\rangle + \sum_{i>2} 
\psi_i \, |\Psi_i \rangle \,,
\ee
where $\Psi_i$, $i>2$, are massive scalars, and the first two fields 
are respectively the tachyon
\be
|T\rangle = c_1 \bar{c}_1 \vac \,,
\ee
and the dilaton (\ref{dilaton-field}). We will often use the 
notation $V_{\psi_{i_1} \psi_{i_2} \ldots \psi_{i_N}}$ to 
designate the coefficient of $\psi_{i_1} \psi_{i_2} \ldots \psi_{i_N}$ in 
the potential (and it is understood that $\psi_1 = t$ and $\psi_2 = d$).

\begin{figure}[!ht]
\begin{center}
\input{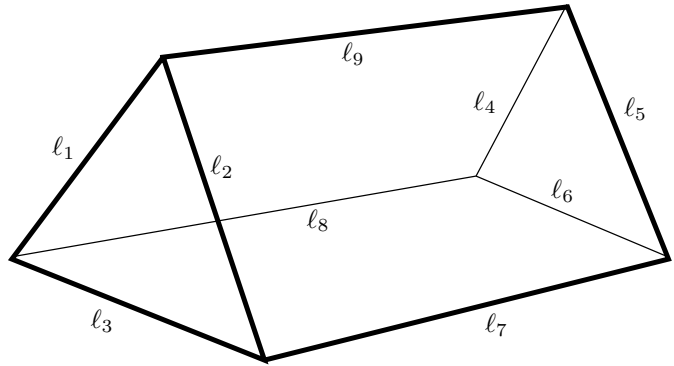}
\caption{\footnotesize{The relevant prism of the five-point
vertex. The labeling of lengths will be kept as shown.}}
\label{prism_f5}
\end{center}
\end{figure}

In the CSFT constructed in \cite{Zwie-BV,CSFT,Saad-Zwie,area}, the
interacting worldsheet id endowed with the following geometry: Every
external state spans a semi-infinite cylinder of perimeter $2 \pi$,
and these cylinders intersect on a prism. The prisms of a contact term
must have the characteristic that all their nontrivial closed curves
must have length greater than or equal to $2 \pi$, which for the
five-point vertex is equivalent to all edges having length smaller
than or equal to $\pi$. It can be shown that all other prisms are
obtained from Feynman diagrams with vertices of lower orders. The
relevant prism for the five-point contact term was first discussed in
\cite{Saad-Zwie}. It is shown on Figure \ref{prism_f5}.
It is made of two opposing triangles, connected with three
quadrilaterals.  The perimeter conditions on the cylinders can be
written
\ben
&& \ell_1 + \ell_2 + \ell_3 = 2 \pi \quad , \quad 
\ell_4 + \ell_5 + \ell_6 = 2 \pi \nonumber \\
&& \ell_7 = \ell_1 + \ell_4 - \pi \quad , \quad 
\ell_8 = \ell_2 + \ell_5 - \pi \quad , \quad 
\ell_9 = \ell_3 + \ell_6 - \pi  \,.
\label{lengths-conditions} \een
Those are five conditions, leaving four independent lengths, matching
the real dimensionality of the moduli space of spheres with five
punctures.  All other prisms with five faces would have less than four
independent lengths and thus correspond to subsets of the moduli space
with measure zero. Although they don't contribute to the integration
over moduli space, some of these prisms will be considered in Section
\ref{Limits_s5} because they will be useful to calculate the boundaries
of the second kind described in Section \ref{ms_s5}.

The usefulness of this particular geometry is that it arises from a
quadratic differential \cite{Strebel} that has second order poles at
the punctures where the external states are inserted, and verifies
the Strebel condition that its critical graph has measure zero. The
ring domains correspond to the semi-infinite cylinders and the
critical graph corresponds to the prism.

To be concrete, we will always map the sphere on the complex plane and
we will fix the topology of the vertex from the beginning, i.e. we
will use the labeling of zeros and poles as indicated on the right of
Figure \ref{mostsym_f5}. Fixing the topology means that we will only
consider configurations obtained from this one by a continuous
transformation and without any two zeros merging.  In other words, the
punctures on quadrilateral faces will always be mapped to $z=0$, $z=1$
and $z=\infty$, whereas the remaining punctures at $\xi_1$ and $\xi_2$
will always correspond to triangular faces. In these notations, the
two complex numbers $\xi_1$ and $\xi_2$ parameterize the moduli space
of five-punctured spheres.

\sectiono{Quadratic differentials}
\label{regular_s5}

To describe the right geometry and the local coordinates 
on a punctured sphere, we need a quadratic differential $\varphi$, 
transforming like 
\be
\varphi = \phi(z) (dz)^2 = \phi(w) (dw)^2
\ee
under a conformal change of variable. It should be holomorphic
everywhere except at the punctures $z_I$, where it has poles of order
two with "residue" minus one, and has thus the expansion
\be
\phi(z) = \frac{-1}{(z-z_I)^2} + {\cal O}\left((z-z_I)^{-1}\right) \,.
\ee
If we place a puncture at infinity, as we will, the expansion of 
$\varphi$ in the coordinate $t=1/z$ is 
\be
\phi(t) = \frac{-1}{t^2} + {\cal O}\left(t^{-1}\right) \,.
\ee
It is easily seen that the quadratic differentials obeying these
conditions are given by
\be
\boxed{\phi(z) = -\frac{z^6 + P_5(z) + a_1 \, z(z-1)(z-\xi_1)(z-\xi_2) + 
a_2 \, z^2(z-1)(z-\xi_1)(z-\xi_2)}
{z^2\, (z-1)^2 \, (z-\xi_1)^2 \, (z-\xi_2)^2}} \,. \label{phi5}
\ee 
Here $P_5(z)$ is a polynomial of order five which is partially
determined by the residue conditions and will be given by
(\ref{h5_2}), and $a_1$ and $a_2$ do not change the residue conditions
at the poles. The coefficient of $z^6$ in the numerator must be $-1$
for the residue at infinity to be $-1$. The parameters $a_1$ and $a_2$
will be determined by the Strebel condition, namely the condition that
the critical graph of $\varphi$ closes (see \cite{Strebel, Belo-Zwie,
Belo, quartic} for more details). It will be enough to know here that
the Strebel solution gives the quadratic differential needed for the
vertex. It can be expressed as the requirement that the complex
lengths between any two zeros $z_1$ and $z_2$ of the quadratic
differential, are real
\be
\Im \int_{z_1}^{z_2} \sqrt{\phi(z)} \, dz = 0 \,.
\ee
Solving numerically the Strebel condition will be the object of
Section \ref{solving_s5}. For now let us go back to $P_5(z)$. We will
write it in the form
\be
P_5(z) = b_5 \, z^5 + b_4 \, z^4 + P_3(z) \,, \label{h5}
\ee
where 
\be
P_3(z) = b_3 \, z^3 + b_2 \, z^2 + b_1 \, z + b_0 \,.
\ee
The four coefficients of $P_3(z)$ can be completely determined, in terms of 
$\xi_1$, $\xi_2$, $b_5$ and $b_4$, by the four residue conditions at the finite poles 
$0$, $1$, $\xi_1$ and $\xi_2$. These conditions can be written
\ben
y_0 &\equiv& P_3(0) = \xi_1^2 \, \xi_2^2 \nonumber \\
y_1 &\equiv& P_3(1) = -1 - b_5 - b_4 + (1-\xi_1)^2 (1-\xi_2)^2 \nonumber \\
y_2 &\equiv& P_3(\xi_1) = -\xi_1^6 - b_5 \, \xi_1^5 - b_4 \, \xi_1^4 + \xi_1^2 (\xi_1 - 1)^2 
(\xi_1 - \xi_2)^2 \nonumber \\
y_3 &\equiv& P_3(\xi_2) = -\xi_2^6 - b_5 \, \xi_2^5 - b_4 \, \xi_2^4 + \xi_2^2 (\xi_2 - 1)^2 
(\xi_1 - \xi_2)^2 \,. \label{hpole}
\een
The polynomial $P_3$ that satisfies (\ref{hpole}) can be written
\ben
P_3(z) &=& \frac{(z-1)(z-\xi_1)(z-\xi_2)}{-\xi_1 \, \xi_2} \, y_0 + 
\frac{z (z - \xi_1) (z - \xi_2)}{(1-\xi_1)(1-\xi_2)} \, y_1 + \nonumber \\
&& + \frac{z(z-1)(z-\xi_2)}{\xi_1(\xi_1-1)(\xi_1-\xi_2)} \, y_2 + 
\frac{z(z-1)(z-\xi_1)}{\xi_2(\xi_2-1)(\xi_2-\xi_1)} \, y_3 \,. \label{Legendre}
\een
Now we want to choose $b_5$ and $b_4$ in (\ref{h5}) in such a way that the expression  
(\ref{Legendre}) is as simple as possible. For this we note, from (\ref{hpole}), 
that if 
\ben
&& b_4 + b_5 \, \xi_1 + \xi_1^2 = 0 \nonumber \\
{\rm and} && b_4 + b_5 \, \xi_2 + \xi_2^2 = 0 \,, \label{b4b5eq}
\een
then the third and fourth terms in (\ref{Legendre}) will 
be simplified. Even better, since the solution of (\ref{b4b5eq}) is 
\ben
b_5 &=& -(\xi_1 + \xi_2) \nonumber \\
b_4 &=& \xi_1 \, \xi_2 \,, \label{b4b5sol}
\een
we have that $-1-b_5-b_4 = - (1-\xi_1)(1-\xi_2)$, and thus we see, from (\ref{hpole}), 
that the second term will be simplified too. 
And obviously, since $y_0 = \xi_1^2 \xi_2^2$, we also have 
that the first term is simplified. Thus, making the choice 
(\ref{b4b5sol}) for $b_5$ and $b_4$, we have that 
\be
\boxed{P_5(z) = -s \, z^5 + t \, z^4 + (-s + v(s-1)) \, z^3 + \left(t+s^2+v(1-s-t) 
\right) \, z^2 + t \left(v - 2s \right) \, z + t^2} \,, \label{h5_2}
\ee
where
\ben
s &\equiv& \xi_1 + \xi_2 \nonumber \\
t &\equiv& \xi_1 \, \xi_2 \nonumber \\
v &\equiv& (\xi_1 - \xi_2)^2 = s^2 - 4 t
\een
are symmetric expressions in $\xi_1$ and $\xi_2$.

\paragraph{}
We note, for future reference, that the derivatives of $\phi(z)$ with 
respect to $a_1$ and $a_2$ are
\ben
\frac{\partial \phi(z)}{\partial a_1} &=& 
\frac{-1}{z \, (z-1) \, (z-\xi_1) \, (z-\xi_2)} \,, \nonumber \\
\frac{\partial \phi(z)}{\partial a_2} &=& 
\frac{-1}{(z-1) \, (z-\xi_1) \, (z-\xi_2)} \,.
\label{dphida} \een

\subsubsection*{The most uniform prism}

\begin{figure}[!ht]
\begin{center}
\input{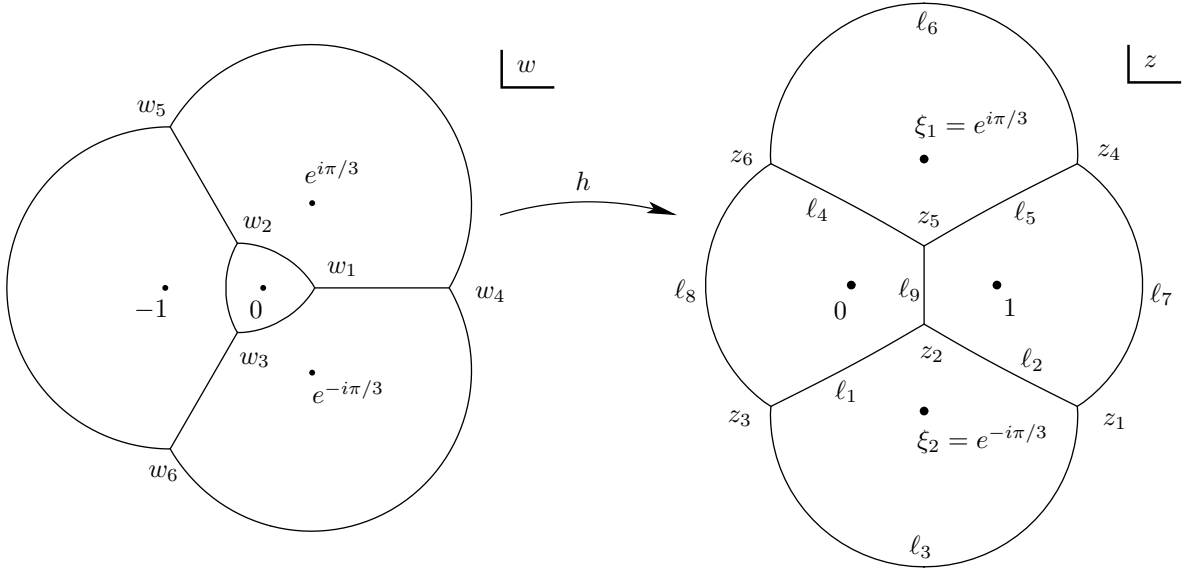}
\caption{\footnotesize{The critical graph of the quadratic
differential corresponding to the most uniform prism, in the
$w$-plane where the symmetry is obvious, and in the $z$-plane in which
the three quadrilateral punctures are at the standard points $0$, $1$,
$\infty$.}}
\label{mostsym_f5}
\end{center}
\end{figure}

We call the {\it most uniform prism}, the prism (Figure
\ref{prism_f5}) which has two equilateral triangles. The lengths of
the edges of the triangles must therefore be $\frac{2 \pi}{3}$, and
the three edges connecting the two triangles must have lengths
$\frac{\pi}{3}$. We want to describe the quadratic differential of
this configuration in the $z$-plane, where we will put three punctures
at respectively zero, one and infinity. As already mentioned, we have
to start by deciding which faces will correspond to these three
punctures. Since we are later going to consider the subgroup of ${\rm
PSL}(2, \mathbb{C})$ conformal transformations that permute the
aforementioned punctures, it will be simplest to map the three
quadrilateral faces on $z=0$, $z=1$ and $z=\infty$. We must then stick
to this choice because it would be very difficult for the numerical
algorithm to switch between configurations of different topologies;
this will be discussed more extensively in Section \ref{ms_s5}.  For
the purpose of finding the quadratic differential of the most uniform
prism, it is easier to work in another coordinate $w$, where the
symmetry of the prism is more visible (see Figure \ref{mostsym_f5}).
In the $w$ coordinate, $\phi(w)$ has three poles at the vertices of an
equilateral triangle, respectively $w=-1$, $w=e^{\frac{i\pi}{3}}$ and
$w=e^{-\frac{i\pi}{3}}$, which correspond to the quadrilateral
faces. One pole is at its center $w=0$, and the last pole is at
infinity.
By contemplating the left half of Figure \ref{mostsym_f5}, we can immediately 
write the ansatz for the zeros $w_i$, $i=1,\ldots,6$, of $\phi(w)$
\ben
w_1 = \beta \,, \quad w_2 = e^\frac{2 i \pi}{3} \beta \,, \quad 
w_3 = e^\frac{4 i \pi}{3} \beta \nonumber \\
w_4 = \gamma \,, \quad w_5 = e^\frac{2 i \pi}{3} \gamma \,, \quad 
w_6 = e^\frac{4 i \pi}{3} \gamma \,,
\een
where $\beta$ and $\gamma$ are positive real numbers and $\gamma > \beta$. 
The quadratic differential for this configuration is thus
\ben
\phi(w) &=& \frac{-(w-\beta) (w-\gamma) (w-e^\frac{2 i \pi}{3} \beta) 
(w-e^\frac{2 i \pi}{3} \gamma) (w-e^\frac{4 i \pi}{3} \beta) 
(w-e^\frac{4 i \pi}{3} \gamma)}
{w^2 \, (w+1)^2 \, \left(w-e^\frac{i \pi}{3}\right)^2 \, 
\left(w-e^{-i\frac{\pi}{3}}\right)^2}
\nonumber \\
&=& -\frac{\left(w^3-\beta^3\right) \left(w^3-\gamma^3\right)}
{w^2 \left(w^3 + 1 \right)^2} \,.
\label{phims}
\een
The residue condition at $w=-1$ and $w=0$ are respectively
\be
\frac{(\beta^3 + 1) \, (\gamma^3+1)}{9} = 1 \qquad , \qquad
\beta^3 \, \gamma^3 = 1   
\,. \label{bg}
\ee
The solution to the system (\ref{bg}) with the constraint 
$\gamma > \beta > 0$ is  
\be
\beta = \left( \frac{7 - 3 \sqrt{5}}{2} \right)^\frac{1}{3} \,, \quad
\gamma = \left( \frac{7 + 3 \sqrt{5}}{2} \right)^\frac{1}{3} \,.
\ee

Now we express the quadratic differential in the $z$ coordinate. We want to 
map the three quadrilateral punctures to $z=0$, $z=1$ and $z=\infty$. Namely
$z = h(w)$, where
\be
h(-1) = 0 \quad , \quad h(e^\frac{i\pi}{3}) = 1 
\quad , \quad h(e^{-\frac{i\pi}{3}}) = \infty \,,
\ee
whence
\be
z = h(w) = e^\frac{i \pi}{3} \frac{1+w}{w-e^{-\frac{i\pi}{3}}} \,,
\ee
which has the inverse
\be
w = h^{-1}(z) = \frac{e^{-\frac{i\pi}{3}} z + e^\frac{i\pi}{3}}{z-e^\frac{i\pi}{3}} \,.
\ee
Remembering that 
\be
\phi(z) = \phi(w) \left(\frac{dw}{dz}\right)^2 \,,
\ee
we find
\be
\phi(z) = -\frac{z^6 - 3 z^5+3z^4-z^3+3z^2-3z+1}{z^2 (z-1)^2 (z-\xi_1)^2 (z-\xi_2)^2} \,,
\label{phi_most_sym} \ee
where the poles $\xi_1$ and $\xi_2$ are
\ben
&& \xi_1 = h(\infty) = e^\frac{i \pi}{3} \nonumber \\
&& \xi_2 = h(0) = e^{-\frac{i\pi}{3}} \,.
\een
Now we can easily determine the parameters $a_1$ and $a_2$ by comparing (\ref{phi5}) 
and (\ref{h5_2}) with (\ref{phi_most_sym}). We find
\be
a_1 = a_2 = -2 \,.
\ee
At last, the zeros of $\phi(z)$ are given by
\be
z_i = h(w_i) \quad , \quad i=1,\ldots,6 \,,
\ee
and their positions are shown on Figure \ref{mostsym_f5}. It is important to 
know an exact quadratic differential for the numerical algorithm to start with. 
It will then cover all the reduced moduli space by successive deformations 
of this solution, each deformation being relatively small for the Newton 
method to converge (more on this in Section \ref{solving_s5}).

\sectiono{Limits of quadratic differentials}
\label{Limits_s5}

One of the most complicated problems in the computation undertaken in
this paper, is to describe the boundary of the reduced moduli
space. As was already noted in \cite{Saad-Zwie}, there are two kinds
of boundaries. When the length of one of the triangle edges is $\pi$,
we are on a boundary of the first kind, corresponding to the situation
in which the vertex can be built as a Feynman diagram with a
propagator of zero length.  There are also boundaries of the second
kind, when one of the lengths $\ell_7$, $\ell_8$ or $\ell_9$ is
zero. These are internal boundaries, they connect prisms with
different assignations of punctures to faces. They are not boundaries
of the whole reduced moduli space, but they are boundaries of the
region ${\cal A}_5$ that we will consider by keeping a fixed
assignment of punctures to faces (see Section \ref{ms_s5}). We will
also need to consider intersections of boundaries, in particular when
two edges have vanishing lengths. The description of the moduli space
will be done in Section \ref{ms_s5}. The problem for now is that when
a length vanishes, the quadratic differential has a double zero and it
becomes numerically ill-conditioned. The goal of this section is
precisely to deal with these limit cases. We can get rid of the
numerical difficulties by some analytical work. Actually in the case
of two double zeros, the Strebel differentials can be explicitly
described in terms of an algebraic curve, reducing the numerical work
to finding the roots of a polynomial of order six.\footnote{Other
analytically solvable limits of quadratic differentials with four and
five poles have been studied in the context of open-closed duality in
\cite{open-closed}.}

\subsection{Quadratic differentials with one double zero}
\label{onedouble_s5}

\subsubsection*{The general case}

When we compute the boundary of the reduced moduli space, we are led
to consider configurations where the quadratic differential has a
double zero, or in other words when an edge collapses to zero
length. We have to distinguish whether the collapsed edge is an edge
of a triangle or one of the three edges connecting the two
triangles. In the first case we end up with a face with only two
edges, and for this configuration to be in the reduced moduli space,
they need to have length $\pi$. The remaining six lengths are
constrained by four residue conditions, and we are therefore left with
two real degrees of freedom. In the second case, we have eight lengths
constrained by five residue conditions, which leave us three degrees
of freedom. We are thus going to consider this case only as this will
describe some boundaries of the reduced moduli space. Once we have
fixed the topology and the labeling of zeros and edges (Figures
\ref{prism_f5} and \ref{mostsym_f5} and
Equ.(\ref{lengths-conditions})) in the $z$ coordinate, we must
describe separately the cases where $\ell_7$, $\ell_8$ or $\ell_9$
respectively, are zero. Since we have now only three real degrees of
freedom we can fix $\xi_1$ and only one real component of $\xi_2$, for
example
\be
(\zeta|\xi_2) \equiv \Re \zeta \Re \xi_2 + \Im \zeta \Im \xi_2 \,,
\ee
where $\zeta$ is a given complex number of unit norm. Once these
quantities are fixed, the quadratic differentials with one double
zero, satisfying the residue conditions have three real degrees of
freedom, which are for example, $\Re u$, $\Im u$ and $(i
\zeta|\xi_2)$, where $u$ is the position of the double zero.

\paragraph{}
We will thus write
\be
\phi(z) = -\frac{(z-u)^2 \left(z^4 + P_3(z)\right)}{z^2 (z-1)^2 (z - \xi_1)^2 (z - \xi_2)^2} \,,
\ee 
where the cubic polynomial 
\be
P_3(z) = c_3 \, z^3 + c_2 \, z^2 + c_1 \, z + c_0 
\ee
will be completely determined by the following residue conditions at the finite poles
\ben
y_0 &\equiv& P_3(0) = \frac{\xi_1^2 \, \xi_2^2}{u^2} \nonumber \\
y_1 &\equiv& P_3(1) = \left( \frac{(1-\xi_1) (1-\xi_2)}{1-u} \right)^2 - 1 \nonumber \\
y_2 &\equiv& P_3(\xi_1) = \left( \frac{\xi_1 (\xi_1-1) (\xi_1-\xi_2)}{\xi_1-u} \right)^2 
- \xi_1^4 \nonumber \\
y_3 &\equiv& P_3(\xi_2) = \left( \frac{\xi_2 (\xi_2-1) (\xi_2-\xi_1)}{\xi_2-u} \right)^2 
- \xi_2^4 \,. \label{h3}
\een
The polynomial $P_3(z)$ that satisfies (\ref{h3}), can be written as in (\ref{Legendre}). 
Now if we define
\ben
v_0 &\equiv& \frac{y_0}{- \xi_1 \, \xi_2} \nonumber \\
v_1 &\equiv& \frac{y_1}{(1-\xi_1) (1-\xi_2)} \nonumber \\
v_2 &\equiv& \frac{y_2}{\xi_1 (\xi_1-1) (\xi_1-\xi_2)} \nonumber \\
v_3 &\equiv& \frac{y_3}{\xi_2 (\xi_2-1) (\xi_2-\xi_1)} \,,
\een
we have that 
\ben
P_3(z) &=& (v_0 + v_1 + v_2 + v_3) \, z^3 -
\left( (1+s) v_0 + s v_1 + (1+\xi_2) v_2 + (1+\xi_1) v_3 \right) \, z^2 + \nonumber \\
&& + \left( (s + t) v_0 + t v_1 + \xi_2 v_2 + \xi_1 v_3 \right) \, z + 
\left(\frac{t}{u} \right)^2 \,,
\een
where again, $s = \xi_1 + \xi_2$ and $t = \xi_1 \, \xi_2$. And the quadratic 
differential is completely determined once we give $\xi_1$, $\xi_2$ and $u$.

\subsubsection*{The regular pyramid}

As we did for the regular configurations, we want to calculate explicitly 
the Strebel quadratic differentials with one double zero in the 
most symmetric case, a pyramid with a square base 
(with edges of length $\frac{\pi}{2}$) and four triangles 
with edges of lengths $\frac{\pi}{2}$, $\frac{3 \pi}{4}$ and 
$\frac{3 \pi}{4}$. Again, it is easier to solve the quadratic differential 
in another coordinate $w$, where the symmetry is obvious, and then map it 
to the $z$-plane. We will have three different mappings $h_1$, 
$h_2$ and $h_3$, whether the 
vanishing length is $\ell_8$, $\ell_9$ or $\ell_7$ respectively.
\begin{figure}[!ht]
\begin{center}
\input{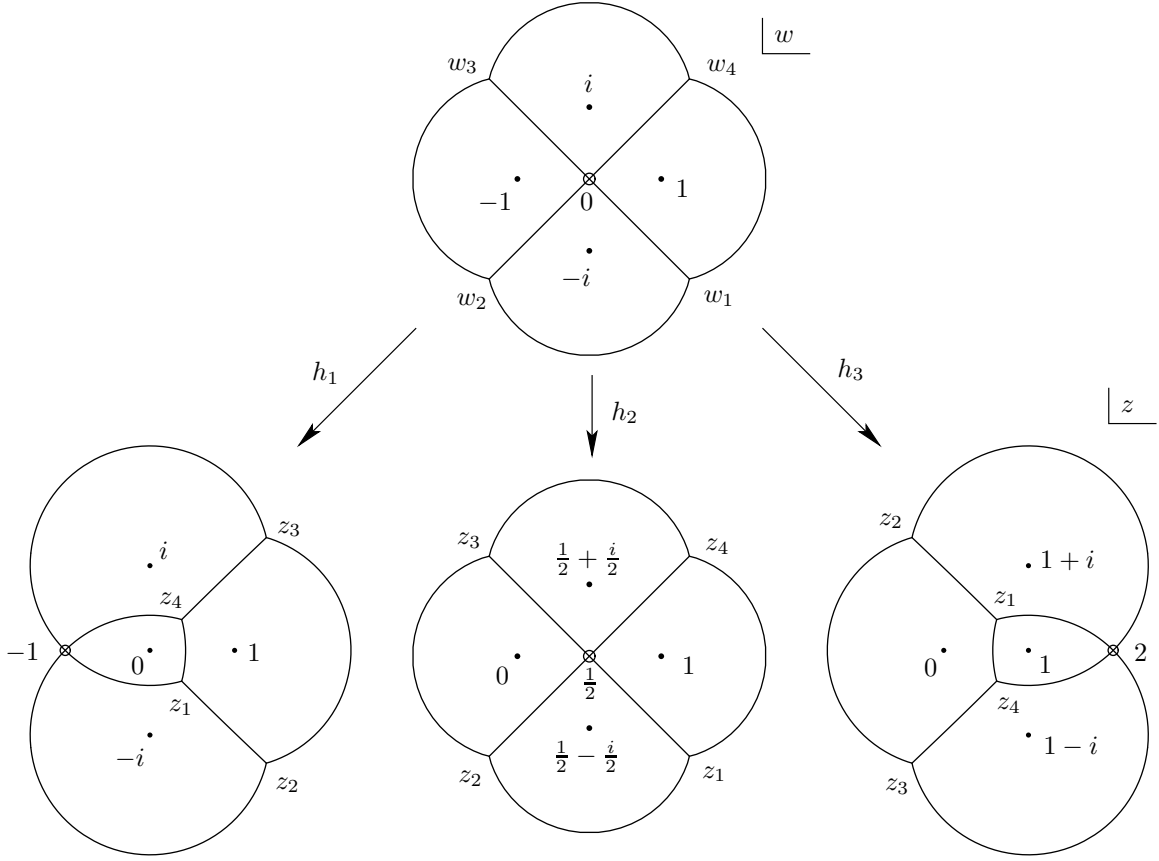}
\caption{\footnotesize{The regular pyramid in the $w$-plane and its
three mappings to the $z$-plane with respectively $\ell_8=0$,
$\ell_9=0$ and $\ell_7=0$. The double zero is marked with a small
circle.}}
\label{pyramid_f5}
\end{center}
\end{figure}
In the $w$ coordinate we set the double zero $u=0$ and four  
poles around it, namely at $1$, $i$, $-1$ and $-i$. The last pole, 
corresponding to the base of the pyramid, is at $w=\infty$. By 
symmetry (see top of Figure \ref{pyramid_f5}), we can then immediately 
read off the ansatz for 
the zeros $w_i$, $i=1\ldots4$ of $\phi(w)$, namely
\be
w_1 = \alpha e^{-\frac{i \pi}{4}} \quad , \quad 
w_2 = \alpha e^{-\frac{3i \pi}{4}} \quad , \quad 
w_3 = \alpha e^\frac{3 i \pi}{4} \quad , \quad 
w_4 = \alpha e^\frac{i \pi}{4} \,,
\ee
where $\alpha$ is a real positive number. We have thus
\be
\phi(w) = -\frac{w^2 (w^4+\alpha^4)}{(w^4-1)^2} \,.
\ee
The residue condition at the pole $w=1$ gives us 
\be
\frac{1+\alpha^4}{16} = 1 \quad \Rightarrow \quad \alpha = (15)^\frac{1}{4} \,.
\ee
We now map it in the $z$-plane to the configuration 
(lower left of Figure \ref{pyramid_f5}) 
which has $\ell_8 = 0$. The map is 
\be
z = h_1(w) = \frac{w-1}{w+1} \,.
\ee
This maps the double zero to 
\be
u^{(1)} = h_1(0) = -1 \,,
\ee
and the poles are
\be
\xi^{(1)}_1 = h_1(i) = i \quad , \quad \xi^{(1)}_2 = h_1(-i) = -i \,.
\ee
And the quadratic differential is 
\be
\phi^{(1)}(z) = - \frac{(z+1)^2 \left(z^4-\frac{7}{2}z^3+6z^2-\frac{7}{2}z+1 \right)}
{z^2 (z-1)^2 (z+i)^2 (z-i)^2} \,.
\ee
For the configuration with $\ell_9=0$ (lower middle of Figure \ref{pyramid_f5}), 
we have
\ben
&& h_2(w) = \frac{1}{2}(w+1) \nonumber \\
&& u^{(2)} = \frac{1}{2} \nonumber \\
&& \xi^{(2)}_1 = \frac{1}{2}(1+i) 
\quad , \quad \xi^{(2)}_2 = \frac{1}{2}(1-i) \nonumber \\
&& \phi^{(2)}(z) = -\frac{\left(z-\frac{1}{2}\right)^2 \left(z^4-2z^3+\frac{3}{2}z^2
-\frac{1}{2}z+1 \right)}{z^2 (z-1)^2 \left(z-\frac{1}{2}(1+i)\right)^2 
\left(z-\frac{1}{2}(1-i)\right)^2} \,.
\een
And for the configuration with $\ell_7=0$ (lower right of Figure \ref{pyramid_f5}), 
we have
\ben
&& h_3(w) = \frac{2}{w+1} \nonumber \\
&& u^{(3)} = 2 \nonumber \\
&& \xi^{(3)}_1 = 1+i 
\quad , \quad \xi^{(3)}_2 = 1-i \nonumber \\
&& \phi^{(3)}(z) = -\frac{(z-2)^2 \left(z^4-\frac{1}{2}z^3+\frac{3}{2}z^2
-2z+1 \right)}{z^2 (z-1)^2 \left(z-(1+i)\right)^2 
\left(z-(1-i)\right)^2} \,.
\een

It is very useful to have these three exact quadratic differentials to
start the Newton method when plotting the boundaries of the second
kind.

\subsection{Quadratic differentials with two double zeros}
\label{twodouble_s5}

As we will see in Section \ref{ms_s5}, it is important to be able to
construct quadratic differentials with two double zeros. In the
present section we will see that we can actually solve the Strebel
condition in terms of an algebraic curve, i.e. everything can be done
algebraically except for a root of a polynomial of order six, which
must be found numerically.

\begin{figure}[!ht]
\begin{center}
\input{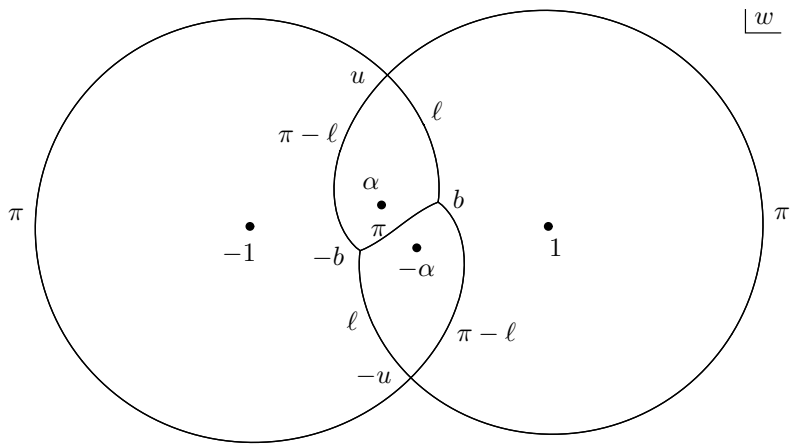}
\caption{\footnotesize{The configuration with two double zeros in the $w$-plane 
with the particular value $\ell=0.8$. The length of each edge is shown.}}
\label{twodouble_f5}
\end{center}
\end{figure}

From Figure \ref{prism_f5} and the lengths conditions
(\ref{lengths-conditions}), we see that when the quadratic
differential has two double zero, one face must have only two sides
(which must therefore have length $\pi$). It can then be further
deduced that only one length is free, with value $\ell$. The other
lengths are respectively $\ell$, $\pi-\ell$ (two edges) and $\pi$. We
now look at the quadratic differential in the $w$-plane where the
puncture at infinity is attached to the face with two edges, and the
punctures of the other two faces which have these edges as a side are
mapped to $w=-1$ and $w=1$. It is readily seen that in this coordinate, 
the critical graph must have the symmetry $w \rightarrow -w$. This is shown on 
Figure \ref{twodouble_f5}.
The two remaining punctures will then be at, say, $w=\alpha$ and
$w=-\alpha$. The two double zeros are at $w=u$ and $w=-u$, and the two
simple zeros are at $w=b$ and $w=-b$. Given the length $\ell$, we want to 
determine the three complex numbers $\alpha$, $u$ and $b$.

The quadratic differential $\varphi = \phi(w)
(dw)^2$ for this configuration is
\be 
\phi(w) = - \frac{(w-u)^2 \,
(w+u)^2 \, (w-b) \, (w+b)} {(w-1)^2 \, (w+1)^2 \, (w-\alpha)^2 \,
(w+\alpha)^2} = - \frac{(w^2-u^2)^2 \, (w^2-b^2)}{(w^2-1)^2 \,
(w^2-\alpha^2)^2} \,.
\label{td5}
\ee
The residue condition at infinity is automatically satisfied by the
expression (\ref{td5}), and by symmetry the residue condition at
$w=-1$ is the same as for $w=1$, and the condition at $w=-\alpha$ is
the same as the one at $w=\alpha$.  We thus have only two independent
residue conditions, at $w=1$ and $w=\alpha$ respectively, which read
\ben
&& \frac{(1-u^2)^2 \, (1-b^2)}{4 \, (1-\alpha^2)^2} = 1 \label{td_equ1} \\
&& \frac{(\alpha^2-u^2)^2 \, (\alpha^2 - b^2)}{4 \, \alpha^2 \, (1-\alpha^2)^2} = 1 
\label{td_equ2} \,. 
\een
We are now going to solve the Strebel condition. For this, we need to write that the complex
lengths between any two zeros are real. By symmetry, this is
automatically satisfied if one length, say the length between $b$ and
$u$, is real. We therefore impose
\be
\ell \equiv \ell(b,u) \in \mathbb{R} \,,
\label{real} \ee
where
\be
\ell(b,u) = \int_b^u \frac{(w^2-u^2) \, \sqrt{b^2-w^2}}{(1-w^2) \, (w^2-\alpha^2)} \, dw 
\,. \label{td_equ3}
\ee
And in total we have two complex equations and one real equation to
determine six real parameters, we thus have one free real parameter
for the configurations with two double zeros. We take this parameter
to be $\ell \in [0,\pi/2]$. The other values $\ell \in [\pi/2, \pi]$
are trivially related to the first case by complex conjugation.

The integral in (\ref{td_equ3}) can be calculated by doing the substitution
$y = \sqrt{1 - \frac{b^2}{w^2}}$, whereby
\ben
&& \int\frac{(w^2-u^2) \, \sqrt{b^2-w^2}}{(1-w^2) \, (w^2-\alpha^2)} \, dw 
= -i \left(\frac{bu}{\alpha} \right)^2 \int
\frac{y^2 \left(y^2 - \left(1-\frac{b^2}{u^2}\right)\right)}
{(y^2-1) \left(y^2-(1-b^2)\right) 
\left(y^2 - \left(1-\frac{b^2}{\alpha^2}\right)\right)} \, dy \nonumber \\
&& = i \int \left( \frac{1}{y^2-1} - 
2 \frac{\sqrt{1-b^2}}{y^2-(1-b^2)} + 
2 \frac{\sqrt{1-\frac{b^2}{\alpha^2}}}{y^2-\left(1-\frac{b^2}{\alpha^2}\right)} \right) \, dy 
\nonumber \\
&& = \frac{i}{2} \ln \left(
\frac{\left( \sqrt{1 - \frac{b^2}{w^2}} - 1 \right) 
\left( \sqrt{1 - \frac{b^2}{w^2}} + \sqrt{1-b^2} \right)^2
\left( \sqrt{1 - \frac{b^2}{w^2}} - \sqrt{1-\frac{b^2}{\alpha^2}} \right)^2}
{\left( \sqrt{1 - \frac{b^2}{w^2}} + 1 \right) 
\left( \sqrt{1 - \frac{b^2}{w^2}} - \sqrt{1-b^2} \right)^2
\left( \sqrt{1 - \frac{b^2}{w^2}} + \sqrt{1-\frac{b^2}{\alpha^2}} \right)^2}
\right) \,. 
\label{td_primitive} \een
To go from the first to the second line we have made use of the residue conditions 
(\ref{td_equ1}) and (\ref{td_equ2}). So the length between $b$ and $u$ is 
\be
\ell = \frac{\pi}{2} + \frac{i}{2} \ln \left(
\frac{\left( \sqrt{1 - \frac{b^2}{u^2}} - 1 \right) 
\left( \sqrt{1 - \frac{b^2}{u^2}} + \sqrt{1-b^2} \right)^2
\left( \sqrt{1 - \frac{b^2}{u^2}} - \sqrt{1-\frac{b^2}{\alpha^2}} \right)^2}
{\left( \sqrt{1 - \frac{b^2}{u^2}} + 1 \right) 
\left( \sqrt{1 - \frac{b^2}{u^2}} - \sqrt{1-b^2} \right)^2
\left( \sqrt{1 - \frac{b^2}{u^2}} + \sqrt{1-\frac{b^2}{\alpha^2}} \right)^2}
\right) \,. \nonumber \\
\ee
It is natural to make the definitions
\be
s \equiv \sqrt{1 - \frac{b^2}{u^2}} \quad , \qquad
t \equiv \sqrt{1 - b^2} \quad , \qquad 
v \equiv \sqrt{1 - \frac{b^2}{\alpha^2}} \,.
\label{stw_def} \ee
We thus have
\be
\frac{(s-1) (s+t)^2 (s-v)^2}{(s+1) (s-t)^2 (s+v)^2} = e^{i (\pi - 2 \ell)} \,.
\label{phase} 
\ee 
We note that the equations (\ref{td_equ1}) and (\ref{td_equ2}) give us a 
simple relation between $t$ and $v$
\be
t - v = - \frac{1}{2} t v \,.
\label{tminusw} \ee
It will be convenient to define
\be
T \equiv t v \,.
\ee
The following identities follow directly from (\ref{tminusw})
\be
t = \sqrt{\frac{T^2}{16} + T} - \frac{T}{4} \quad , \qquad 
v = \sqrt{\frac{T^2}{16} + T} + \frac{T}{4} \,.
\label{tw_T} \ee
We also have from (\ref{td_equ1}), (\ref{stw_def}) and (\ref{tw_T})
\be
s^2 = -\frac{3 T (T-4)}{4 (1+T)} \,,
\label{s2} \ee
and 
\be
u = \frac{ib}{\sqrt{s^2-1}} \quad , \qquad \alpha = \frac{ib}{\sqrt{v^2-1}} \,.
\label{ua_T} \ee
Let us now rewrite (\ref{phase}) as a polynomial equation in $s$,
whose coefficients are expressed in terms of $T$ and $\ell$ with the help 
of (\ref{tw_T})
\be
-i \cot(\ell) \, s^5 - (1+T) s^4  - 
i \cot(\ell) \, \left(\frac{T^2}{4} - T \right) s^3 +
\left( \frac{3}{4} T^2 + 2 T \right) - T^2 = 0 \,.
\ee
Using (\ref{s2}) this becomes
\be
s = -\frac{i}{3} \tan(\ell) \, 
\frac{18 (T+1) \left(T - \frac{16}{9}\right)}{(T-4)^2} 
\ee
Squaring and using again (\ref{s2}), we finally get a polynomial equation 
for $T$.
\be
P_\ell(T) \equiv -\cot^2(\ell) \, T (T-4)^5 + 48 (T+1)^3 \left(T-\frac{16}{9}\right)^2 = 0\,.
\label{T_eq} \ee
We immediately see the solutions of this equation for the particular values 
$\ell = \pi/2$ and $\ell \rightarrow 0$. Namely $T = 16/9$ and $T=4$ 
respectively. For generic $\ell$, this equation must be solved numerically, 
following the branch $T(\ell = \pi/2) = 16/9$.

\begin{figure}[!ht]
\begin{center}
\input{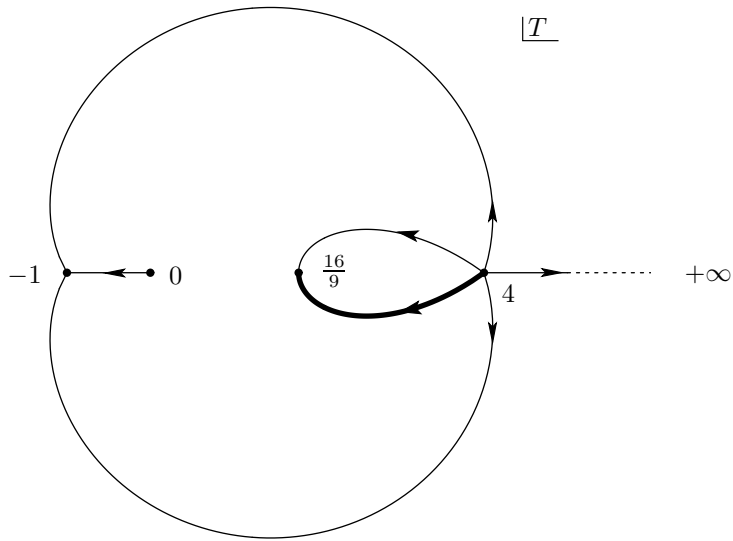}
\caption{\footnotesize{The branch structure of Equ.(\ref{T_eq}). The branch chosen 
is the bold one.}}
\label{branches_f5}
\end{center}
\end{figure}

\paragraph{}
We would like to characterize precisely the branch of the solution of 
(\ref{T_eq}) that we must choose. We start by showing that the branches 
cross only at the points $T=\frac{16}{9}$ and $T=4$. Branches cross at 
multiple zeros, when $P_\ell(T)$ and $P_\ell'(T)$ vanish simultaneously.
We have
\be
P_\ell'(T) = (3T-2) \left( -2 \, \cot^2(\ell) (T-4)^4 + \frac{5}{3} \, 
48 (T+1)^2 \left(T-\frac{16}{9} \right) \right) \,.
\ee
We define
\be
A \equiv -\cot^2(\ell) (T-4)^4 \qquad , \qquad B \equiv 48 (T+1)^2 \left(
T-\frac{16}{9} \right) \,,
\ee
so that we have the system
\ben
0 &=& P_\ell(T) = T(T-4) \, A + (T+1) \left(T-\frac{16}{9}\right) \, B \\
0 &=& P_\ell'(T) = (3 T-2) \left(2 \, A + \frac{5}{3} \, B \right) \,.
\een
Let us look at the second equation. If $T=\frac{2}{3}$, (\ref{T_eq})
tells us that $\cot^2(\ell) =-1$. If $T\neq\frac{2}{3}$, we must have
$B = -\frac{6}{5} A$, and the first equation tells us that either
$T=\frac{2}{3}$ or $T=-16$. In both of these cases we see from
(\ref{T_eq}), that $\cot^2(\ell) = -1$. We have thus shown that the
branches do not cross when $\ell \in (0, \frac{\pi}{2})$. From this
fact we can completely derive the topology of the branch diagram (by
which we mean the locus in the $T$-plane formed by all roots of
$P_\ell(T)$ for all $\ell \in [0,\frac{\pi}{2}]$).  Since the
polynomial $P_\ell(T)$ is real, the branch diagram must be symmetric
under complex conjugation. Five branches must start (at $\ell=0$) from
$T=4$ and one must start from $T=0$, whereas three branches end (when
$\ell = \frac{\pi}{2}$) at $T=-1$ and two must end at $T =
\frac{16}{9}$. Also when $\ell \rightarrow \frac{\pi}{2}$,
$\cot^2(\ell)$ tends to zero and we have a solution
$$
T \approx \frac{48}{\cot^2(\ell)} \rightarrow +\infty \,,
$$ 
and therefore one branch must end at $+\infty$. It is then clear
that the only topology compatible with all these observations is the
one shown on Figure \ref{branches_f5}.
The branch corresponding to the conventions of Figure \ref{twodouble_f5} is 
the one drawn with a bold line. It can be characterized by the fact that it 
belongs to the rectangle $\frac{16}{9} \leq \Re T \leq 4$ and 
$-0.6 \leq \Im T \leq 0$.
We will thus define the function $T(\ell)$ by 
\be
\boxed{T(\ell) \ {\rm is \ the \ unique \ solution \ of} \  \left\{ 
\begin{array}{c}
\displaystyle{-\cot^2(\ell) \, T (T-4)^5 + 48 (T+1)^3 \left(T-\frac{16}{9}\right)^2 = 0} \crbig
\displaystyle{\frac{16}{9} \leq \Re T \leq 4} \crbig
\displaystyle{-0.6 \leq \Im T \leq 0} \end{array} \right.}
\label{Tell}
\ee
From (\ref{stw_def}), (\ref{tw_T}) and (\ref{ua_T}), we can write the expressions of 
$b(\ell)$, $u(\ell)$ and $\alpha(\ell)$ in terms of $T(\ell)$. The only one we 
will need is $\alpha(\ell)$
\be
\boxed{\alpha(\ell) = i \, \frac{1-\frac{T(\ell)^2}{8}-T(\ell) + 
\frac{T(\ell)}{2} 
\sqrt{\frac{T(\ell)^2}{16} + T(\ell)}}
{\sqrt{2T(\ell)-1-\frac{3}{4} 
T(\ell)^2}}} \,.
\label{alphal} \ee
This expression will be useful later do describe the boundary of the projection 
of the reduced moduli space on the $\xi_1$-plane. We already note the 
special values
\be
\alpha(0) = 2-\sqrt{5} \quad , \qquad \alpha(\pi / 2) = 
\frac{27}{32 \sqrt{6} + 19 \sqrt{15}} \, i \,.
\label{alphabounds} \ee

\sectiono{Solving a quadratic differential numerically}
\label{solving_s5}

We now want to describe the numerical algorithm to solve a quadratic
differential. We will focus on the regular quadratic differentials,
but the techniques can be applied, with only trivial modifications, to
quadratic differentials with one double zero. We will then recall how
to compute the mapping radii, and we'll shortly discuss how we compute
derivatives with respect to $\xi_i$ and $\bar{\xi_i}$.

\paragraph{}
The central problem is to find, for given poles $\xi_1$ and $\xi_2$, 
the quadratic differential (i.e. its parameters $a_1$ and $a_2$) that satisfies 
the Strebel condition. In other words its critical graph must close, with 
its zeros being linked with horizontal trajectories. We remind that horizontal 
trajectories are defined by the condition $\phi(z) (dz)^2 > 0$, so the 
Strebel condition is equivalent to the condition that all the complex lengths 
between zeros, $z_i$ and $z_f$, are real.
\be
\Im \ell(z_i, z_f) = \Im \int_{z_i}^{z_f}\sqrt{\phi(z)} \, dz = 0 
\quad , \quad i, f = 1,\ldots,6 \,. 
\ee
We have nine lengths (see Figure \ref{mostsym_f5}) constrained by the five 
residue conditions; this leaves us four independent lengths, which we can 
choose as $\ell_6$, $\ell_4$, $\ell_9$ and $\ell_2$ 
(as labeled on Figure \ref{mostsym_f5}). We thus have to solve a system of 
four real equations of four real unknowns 
\be
{\bf f}({\bf x}) = 0 \,,
\ee
with ${\bf f} = (\Im\ell_6, \Im\ell_4, \Im\ell_9, \Im\ell_2)^T$ and 
${\bf x} = (\Re a_1, \Im a_1, \Re a_2, \Im a_2)^T$. We use Newton's method
\be
{\bf x}_{i+1} = {\bf x}_i - \left(\frac{\partial {\bf f}}{\partial {\bf x}} 
\right)^{-1} \cdot {\bf f}({\bf x}_i) \,,
\ee
which converges when the initial guess ${\bf x}_0$ is not too far from 
the solution. The Jacobian $\partial {\bf f} / \partial {\bf x}$ can be 
expressed in terms of derivatives with respect to $a_i$
\be
\frac{\partial {\bf f}}{\partial {\bf x}} = \left( \begin{array}{cccc}
\Im \partial_{a_1}\ell_6 & \Re \partial_{a_1}\ell_6 & 
\Im \partial_{a_2}\ell_6 & \Re \partial_{a_2}\ell_6 \\
\Im \partial_{a_1}\ell_4 & \Re \partial_{a_1}\ell_4 & 
\Im \partial_{a_2}\ell_4 & \Re \partial_{a_2}\ell_4 \\
\Im \partial_{a_1}\ell_9 & \Re \partial_{a_1}\ell_9 & 
\Im \partial_{a_2}\ell_9 & \Re \partial_{a_2}\ell_9 \\
\Im \partial_{a_1}\ell_2 & \Re \partial_{a_1}\ell_2 & 
\Im \partial_{a_2}\ell_2 & \Re \partial_{a_2}\ell_2
\end{array} \right)
\ee 

\paragraph{}
We continue this section by explaining how to accurately compute numerically 
the complex lengths and their derivatives with respect to $a_i$. And we will 
continue by describing the computation of the mapping radii, and of the 
derivatives of $a_i$ with respect to $\xi_j$ and $\bar{\xi_j}$.

\subsection{The complex lengths}
\label{lengths_s5}

The quintic numerical computation turns out to be very much harder
than the quartic. In the quartic computation \cite{quartic}, there
were two weaknesses which had negligible consequences. First it was
hard to tell if the path of integration (which was chosen to be a
straight line) was going on the right side of the poles, hence a
possible ambiguity of $2 \pi \, n$ in the length. Combined with the
sign ambiguity of the square root, there was a potential problem when
the length was nearly $\pi$. Although we could go around this
ambiguity in the quartic case, for the quintic calculation it would be
catastrophic. Second, we used the coded "continuous square root",
which remembers its last evaluation and tries to detect if the branch
cut has been crossed between the last two evaluations. Again this is
not good enough in the quintic case; we would inevitably encounter
situations when the continuous square root fails to detect a branch
cut, again with catastrophic consequences.

\begin{figure}[!ht]
\begin{center}
\input{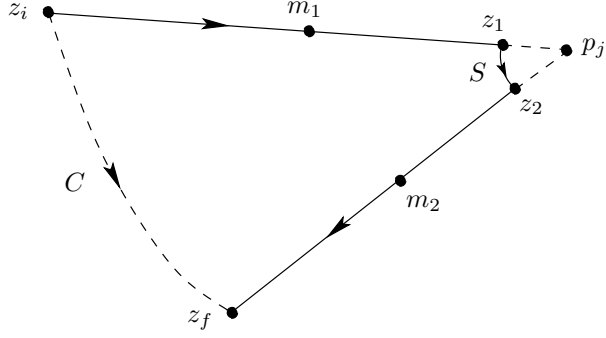}
\caption{\footnotesize{The critical trajectory $C$ can always be deformed 
continuously to the contour integration shown with solid lines.}}
\label{int_path_f5}
\end{center}
\end{figure}

To solve the first problem, we make a conjecture based on observation
that the integration path between two zeros $z_i$ and $z_f$ along the
critical trajectory $C$ that connects them, can be continuously
deformed to the path shown on Figure \ref{int_path_f5}, $z_i$ to $p_j$
to $z_f$ without crossing any pole (other than $p_j$). Here $p_j$ is a
pole such that the zeros $z_i$ and $z_f$ are on the boundary of its
ring domain.
\be
\ell(z_i, z_f) = \int_{C}\sqrt{\phi(z)} \, dz \nonumber
= \int_{z_i}^{z_1}\sqrt{\phi(z)}\,dz + 
\int_S\sqrt{\phi(z)}\,dz + \int_{z_2}^{z_f}\sqrt{\phi(z)}\,dz \,.
\ee
For numerical reasons that will become clear in a while, we split the 
integrals along the straight lines in two, by setting 
$m_1 = \frac{1}{2} (z_i+p_j)$ and $m_2 = \frac{1}{2} (z_f+p_j)$, and 
writing
\be
\ell(z_i,z_f) = \int_{z_i}^{m_1}\sqrt{\phi(z)}\,dz + \int_{m_1}^{z_1}\sqrt{\phi(z)}\,dz + 
\int_S\sqrt{\phi(z)}\,dz + \int_{z_2}^{m_2}\sqrt{\phi(z)}\,dz +
\int_{m_2}^{z_f}\sqrt{\phi(z)}\,dz \,.
\label{lint_split} \ee
Now we subtract the poles at $z=p_j$ in the integrals from $m_k$ to $z_k$, and 
put them back together with the integral over $S$, and we take the limit
$z_1, z_2 \rightarrow p_j$.
\ben
\ell(z_i,z_f) &=& \int_{z_i}^{m_1}\sqrt{\phi(z)}\,dz - \int_{z_f}^{m_2}\sqrt{\phi(z)}\,dz 
\nonumber \\
&& + \int_{m_1}^{p_j}\left(\sqrt{\phi(z)} - \frac{r}{z-p_j} \right)\,dz - 
\int_{m_2}^{p_j}\left(\sqrt{\phi(z)} - \frac{r}{z-p_j} \right)\,dz 
\nonumber \\
&& + \lim_{\genfrac{}{}{0pt}{}{z_1, z_2 \rightarrow p_j}{|z_1-p_j| = |z_2-p_j|}} 
\left( \int_S\sqrt{\phi(z)}\,dz + 
r \int_{m_1}^{z_1}\frac{1}{z-p_j}\,dz - r \int_{m_2}^{z_2}\frac{1}{z-p_j}\,dz
\right) \,,
\label{lint} \een 
where $r = \pm i$ is the residue of $\sqrt{\phi(z)}$ at $z = p_j$.
It is now time to define $\sqrt{\phi(z)}$. 
We want it to be continuous on the integration path, in other 
words, we don't want to cross any branch cut. The idea is to calculate the intersection 
of the integration path with the branch cut of the conventional square root, and then 
change the sign according to the number $n(z)$ of branch points that were crossed. 
Since $\phi(z)$ is given by an expression of the form
\be
\phi(z) = \frac{-P(z)}{D(z)^2} \,,
\ee
we write
\be
\sqrt{\phi(z)} \equiv \frac{\sqrt{-\eta^2 P(z)}}{\eta D(z)} (-1)^{n(z)} \,,
\label{sqrt_def} \ee
where the square root on the right hand side is the conventional square root 
with branch cut the negative real axis.
We ignore $\eta$ for now and focus on $n(z)$. When $z_0$ is on the integration path, 
say $[z_i,p_j]$, we define $n(z_0)$ to be the number 
of times that the conventional square root 
$\sqrt{-\eta^2 P(y)}$ crosses its branch cut when $y \in [z_0, p_j]$, 
In particular $n(p_j) = 0$. We start by parameterizing the path with 
$t \in [-1,1]$
\be
z(t) = \frac{p_j-z_i}{2} \, t + \frac{p_j+z_i}{2} \equiv az+b \,,
\ee
$n(z_0)$ is then given by the number of solutions $t_i$
of the equation
\be
\Im\left(-\eta^2 P(z(t_i)) \right) = 0 \,,
\label{branch_cut} \ee
satisfying the conditions 
\be
\Re\left(-\eta^2 P(z(t_i)) \right) < 0 \qquad {\rm and} \qquad  
t_0 < t_i < 1 \,,
\ee
where $z(t_0) = z_0$. 
It is then clear that (\ref{sqrt_def}) will be continuous. Let us now look at the 
complex number $\eta$ in (\ref{sqrt_def}). We want to choose it in such a way 
that the residue of $\sqrt{\phi(z)}$ at $z = p_j$, which is $\pm i$, takes the 
negative sign. Remembering that the ``residues'' of $\phi(z)$ at the poles are $-1$, 
we see that $P(p_j) = \prod_{i\neq j}(p_j-p_i)^2 = (D'(p_j))^2$. 
We thus require that $\eta$ satisfies
\be
\frac{\sqrt{-\left(\eta D'(p_j) \right)^2}}{\eta D'(p_j)} = -i \,,
\ee
which is the case if 
\be
\Im\left(\eta D'(p_j)\right) > 0 \qquad {\rm or} \qquad 
\left( \Im\left(\eta D'(p_j)\right) = 0 \quad {\rm and} \quad 
\Re\left(\eta D'(p_j)\right) < 0 \right) \,.
\ee
We will therefore take (demanding also that $\eta$ has unit norm)
\be
\eta = i \, \frac{\left|D'(p_j)\right|}{D'(p_j)} \, e^{i \theta} \,,
\label{eta-def} \ee
where $\theta \in (-\frac{\pi}{2}, \frac{\pi}{2})$ can be chosen arbitrarily, 
as long as as it is not zero. Indeed if $\theta = 0$, we would have 
$-\eta^2 P(p_j) > 0$, and (\ref{branch_cut}) would have the solution 
$t=1$. This must be avoided because, due to numerical uncertainties, we would 
find a solution $t = 1 \pm \epsilon$ and it would be hard to decide whether to 
count it in $n(z)$ or not. If $\theta \neq 0$, we avoid this problem.

We can now continue the computation (\ref{lint}) by setting $r=-i$. In the 
limit considered, the integral over the arc of circle $S$ is just a bit of the 
residue, namely 
\be
\int_S\sqrt{\phi(z)}dz = \alpha i (-i) = \alpha\,,
\ee
where $\alpha$ is the angle at the tip $p_j$ of the triangle 
$z_ip_jz_f$. We thus have
\be
\int_S\sqrt{\phi(z)}dz = \arg^* \frac{z_f-p_j}{z_i-p_j} \,,
\label{Sint} \ee
where $\arg^*$ is the argument function with range $[0,2\pi)$, 
in other words, its branch cut is on the 
positive real axis. This is necessary because $\alpha$ might be greater 
than $\pi$, when $p_j$ is on the right of the straight line from 
$z_i$ to $z_f$ but still on the left of the path $C$. The last 
two integrals in (\ref{lint}) give
\ben
&& \lim_{\genfrac{}{}{0pt}{}{z_1, z_2 \rightarrow p_j}{|z_1-p_j| = |z_2-p_j|}} 
\left(
-i \int_{m_1}^{z_1}\frac{1}{z-p_j}\,dz + i \int_{m_2}^{z_2}\frac{1}{z-p_j}\,dz
\right) 
\nonumber \\
&& = -i \lim_{\genfrac{}{}{0pt}{}{z_1, z_2 \rightarrow p_j}{|z_1-p_j| = |z_2-p_j|}} 
\log \left(\frac{z_1-p_j}{z_2-p_j} 
\frac{z_f-p_j}{z_i-p_j} \right) 
= -i \log \left|\frac{z_f-p_j}{z_i-p_j} \right| \,,
\label{logint} \een
where in the last equality, we used the fact that $z_1$ is on the segment 
from $z_i$ to $p_j$ and $z_2$ is on the segment 
from $z_f$ to $p_j$, and the argument of the log is therefore a 
positive real number. Combining (\ref{Sint}) and (\ref{logint}), we see 
that the last line of (\ref{lint}) is 
$$
-i \log^* \frac{z_f-p_j}{z_i-p_j} \,,
$$
where $\log^*z \equiv \log|z| + i \arg^*z$. In total we 
have therefore
\ben
\ell(z_i,z_f) &=& \int_{z_i}^{m_1}\sqrt{\phi(z)}\,dz + 
\int_{m_1}^{p_j}\left(\sqrt{\phi(z)} + \frac{i}{z-p_j} \right)\,dz
\nonumber \\
&& - \left( \int_{z_f}^{m_2}\sqrt{\phi(z)}\,dz + 
\int_{m_2}^{p_j}\left(\sqrt{\phi(z)} + \frac{i}{z-p_j} \right)\,dz
\right) - i \, \log^*\frac{z_f-p_j}{z_i-p_j} \,.
\label{lint2} \een
We shall now explain why we have split in two the integrals along 
the straight lines. For this we need to make a parenthesis into the 
Gaussian quadrature formulas (see for example \cite{Rals-Rabi}) 
which we use to compute numerically the 
integrals in (\ref{lint2}). Those are very useful when integrating, 
over the finite interval $[-1,1]$, a function $g(t)$ that  behaves like 
$(1-t)^\alpha$ near $t=1$, and like $(1+t)^\beta$ near $t=-1$, 
where $\alpha, \beta > -1$. If $\alpha$ or $\beta$ are not integer 
the function $g(t)$ will have singular derivatives, and a quadrature 
formula like the trapezoidal method or the Gauss method with no weight 
(which is adapted when $\alpha = \beta = 0$) will give inaccurate 
results. In that case, we need to use the Gauss-Jacobi quadrature 
formula
\be
\int_{-1}^1(1-t)^\alpha(1+t)^\beta f(t) dt = \sum_{j=1}^N H_j f(a_j) + E \,,
\ee
where 
\be
g(t) = (1-t)^\alpha(1+t)^\beta f(t) \,,
\label{gf} \ee
and $f(t)$ is infinitely 
differentiable with all its derivatives bounded on $[-1,1]$. The 
abscissas $a_j$ are the roots of the Jacobi polynomial $J_N(t; \alpha,\beta)$, 
the weights $H_j$ are given by 
\be
H_j = - \frac{2N+\alpha+\beta+2}{N+\alpha+\beta+1} \,
\frac{\Gamma(N+\alpha+1) \Gamma(N+\beta+1)}
{\Gamma(N+\alpha+\beta+1) (N+1)!} \, 
\frac{2^{\alpha+\beta}}{J'_N(a_j; \alpha, \beta) 
J_{N+1}(a_j; \alpha, \beta)} \,,
\ee
and the error $E$ is
\be
E = \frac{\Gamma(N+\alpha+1) \Gamma(N+\beta+1) \Gamma(N+\alpha+\beta+1)}
{(2N+\alpha+\beta+1) \left( \Gamma(2N+\alpha+\beta+1) \right)^2} \, 
\frac{N! 2^{2N+\alpha+\beta+1}}{(2N)!} \, f^{(2N)}(\zeta) \quad ,
\quad \zeta \in (-1,1).
\ee
We take $N=15$ for all our integrations, which is enough to obtain a
typical accuracy of fourteen significant digits.

Now suppose that we didn't split the integrals in (\ref{lint_split}) 
and tried to integrate
$$
\int_{-1}^1 \left( \sqrt{\phi(z(t))} + \frac{i}{z(t)-p_j} \right) dt \,.
$$
At $t=-1$, the first term behave like $(t+1)^{1/2}$, but because of the 
second term, we cannot write the integrand in the form (\ref{gf}). 
This is why we had to split the integrals in (\ref{lint_split}). 
In this way, the first and third integrals in (\ref{lint2}) are of 
the form (\ref{gf}) with $\alpha=0$ and $\beta=\frac{1}{2}$, while 
the second and fourth integrals have $\alpha=\beta=0$. 

Note that there is an alternative to the splitting, which was used in 
\cite{quartic}. It is to subtract to $\sqrt{\phi(z)}$ the expression
$$
\frac{-i}{\sqrt{2}} \, \frac{\sqrt{t+1}}{(z(t) - p_j)}
$$
instead of $-i / (z(t) - p_j)$. In that way we can factor out 
the square root along the whole path. But we find the splitting method 
easier, because the expression 
$$
\sqrt{\phi(z)} + \frac{i}{z-p_j}
$$
is simpler when one simplifies it by explicitly canceling the poles. 
What we mean is that this expression is prone to numerical errors, because 
when $z$ is close to $p_j$ we are subtracting two large and almost equal 
numbers, a cancellation error. So we need to manipulate this expression. 
Namely
\ben
\sqrt{\phi(z)} + \frac{i}{z-p_j} &=& 
\frac{\sqrt{-\eta^2 P(z)}}{\eta D(z)} + \frac{i}{z-p_j} 
= \frac{\sqrt{-\eta^2 P(z)} + i \eta D_j(z)}{\eta D(z)}
\nonumber \\
&=& \eta \frac{D_j(z)^2-P(z)}{z-p_j} \, 
\left( D_j(z) \left( \sqrt{-\eta^2 P(z)} - i \eta D_j(z) \right) 
\right)^{-1} \,,
\label{cancellation} \een
where we have defined 
\ben
D_j(z) \equiv \frac{D(z)}{z-p_j} = \prod_{i\neq j}(z-p_i) \,.
\een
We can cancel the apparent pole in the factor 
\be
Q(z) \equiv \frac{D_j(z)^2-P(z)}{z-p_j}
\label{Q} \ee
because the residue condition of the quadratic differential at its poles 
implies that $P(p_j) = D_j(p_j)^2$, so the numerator of (\ref{Q}) is a 
polynomial with a factor $z-p_j$. Thus $Q(z)$ is a polynomial, and the 
right hand side of (\ref{cancellation}) is well defined, analytically as 
well as numerically.

\paragraph{}
In order to use the Newton method, we still need to calculate the 
derivatives $\partial_{a_j} \ell(z_i, z_f)$. Since $\phi(z)$ vanishes 
at the endpoints of the integration path, we can simply differentiate 
inside the integral
\be
\frac{\partial}{\partial a_j} \ell(z_i, z_f) = 
\frac{\partial}{\partial a_j} \int_C \sqrt{\phi(z)} \, dz 
= \frac{1}{2} \int_C \frac{1}{\sqrt{\phi(z)}} 
\frac{\partial \phi(z)}{\partial a_j} \, dz \,.
\ee
With the expressions (\ref{dphida}) for the derivatives and the definition 
(\ref{sqrt_def}) of the square root, and deforming the integration contour 
as before we find
\ben
\frac{\partial}{\partial a_1} \ell(z_i, z_f) &=&
-\frac{\eta}{2} \left( \int_{z_i}^{p_j}\frac{(-1)^{n(z)}}{\sqrt{-\eta^2 P(z)}} \, dz - 
\int_{z_f}^{p_j}\frac{(-1)^{n(z)}}{\sqrt{-\eta^2 P(z)}} \, dz \right) \nonumber \\
\frac{\partial}{\partial a_2} \ell(z_i, z_f) &=&
-\frac{\eta}{2} \left( \int_{z_i}^{p_j}\frac{z \, (-1)^{n(z)}}{\sqrt{-\eta^2 P(z)}} \, dz - 
\int_{z_f}^{p_j}\frac{z \, (-1)^{n(z)}}{\sqrt{-\eta^2 P(z)}} \, dz \right) \,.
\een
These integrals can be evaluated accurately with a Gauss-Jacobi formula 
with $\alpha = 0$ and $\beta = -\frac{1}{2}$.

\subsection{The mapping radii}
\label{radii_s5}

We recall that the mapping radii $\rho_j$ appear as a multiplicative
factor in front of the local coordinate $w_j$ in the power expansion
of the maps $h_j$ from the local coordinates to the uniformizer $z$
(or $t=1/z$ for the puncture at infinity) (see
Equ.(\ref{h-expand})). While all other terms in the expansion can be
determined by expanding the quadratic differential, the mapping radii
can't. Instead, they must be computed by integrating $\sqrt{\phi(z)}$
from a point on the boundary of the ring domain of the pole, to the
pole $p_j$, and subtracting the singular logarithm (see
\cite{Belo-Zwie, quartic})
\be 
\log \rho_j = \lim_{\epsilon \rightarrow 0} \left(
\Im\int_{z_i}^{p_j + \epsilon \alpha} \sqrt{\phi(z)} dz + \log
\epsilon \right) \quad , \quad \alpha = \frac{z_i-p_j}{|z_i-p_j|} \,,
\ee 
and we take $z_i$ to be one of the zeros on the boundary of the
ring domain.  The square root is again defined as in (\ref{sqrt_def})
with $\eta$ as in (\ref{eta-def}), so the residue of $\sqrt{\phi(z)}$
at $z=p_j$ is $-i$. This integral is very similar to that for the
complex lengths, so we know how to evaluate it numerically. First we
break the integration in two and regularize the integrand near the
pole
$$
\log \rho_j = \lim_{\epsilon \rightarrow 0} \left(
\Im \int_{z_i}^{z_1}\sqrt{\phi(z)} dz + \Im \int_{z1}^{p_j+\epsilon \alpha} 
\left(\sqrt{\phi(z)} + \frac{i}{z-p_j} \right) dz + \log \epsilon  - 
\Re \int_{z_1}^{p_j+\epsilon \alpha} \frac{1}{z-p_j} dz \right)
$$
where for definiteness we take $z_1 = \frac{1}{2} (z_i+p_j)$. Now we can 
easily get rid of the limit
\ben
\log \rho_j &=& \lim_{\epsilon \rightarrow 0} \left(
\Im \int_{z_i}^{z_1} \sqrt{\phi(z)} dz + \Im \int_{z_1}^{p_j+\epsilon \alpha} 
\left( \sqrt{\phi(z)} + \frac{i}{z-p_j} \right) dz - \log \left|\frac{\epsilon \alpha}
{z_1-p_j} \right| + \log \epsilon \right) \nonumber \\
&=& \Im \int_{z_i}^{z_1} \sqrt{\phi(z)} dz + \Im \int_{z_1}^{p_j} \left(
\sqrt{\phi(z)} + \frac{i}{z-p_j} \right) dz + \log \left|\frac{z_i-p_j}{2} 
\right| \,.
\een
The integrand of the second term is manipulated as in (\ref{cancellation}) - 
(\ref{Q}), and both integrals can be evaluated accurately with Gauss-Jacobi 
quadrature formulas.

\paragraph{}
The pole at infinity must be treated separately; 
we must work in the coordinate $t=1/z$. The radius is then
\be
\log \rho_5 = \lim_{\epsilon \rightarrow 0} \left(
\Im \int_{1/z_i}^{\epsilon \alpha} \sqrt{\phi(t)} dt + \log \epsilon \right) \,,
\ee
where $z_i$ is a zero on the boundary of the ring domain at infinity. One could 
of course rewrite the integral in the $z$ coordinate
$$
\int_{1/z_i}^{\epsilon \alpha} \sqrt{\phi(t)} dt = 
\int_{z_i}^\frac{1}{\epsilon \alpha} \sqrt{\phi(z)} dz \,,
$$
but we prefer to integrate numerically over a finite interval, so we will stay 
in the $t$ coordinate. We have thus
\ben
\log \rho_5 &=& \lim_{\epsilon \rightarrow 0} \left(
\Im \int_\frac{1}{z_i}^\frac{1}{2z_i}\sqrt{\phi(t)} dt + \Im 
\int_\frac{1}{2zi}^{\epsilon \alpha} 
\left(\sqrt{\phi(t)} + \frac{i}{t} \right) dt + \log \epsilon  - 
\Re \int_\frac{1}{2 z_i}^{\epsilon \alpha} \frac{1}{t} dt \right)
\nonumber \\
&=& \Im \int_\frac{1}{z_i}^\frac{1}{2z_i} \sqrt{\phi(t)} dt + 
\Im \int_\frac{1}{z_i}^0 \left(
\sqrt{\phi(t)} + \frac{i}{t} \right) dt + \log \left|\frac{1}{2 z_i} 
\right| \,.
\een
The integrand of the second term can be manipulated again, as in 
(\ref{cancellation}) - (\ref{Q}), in order to get rid of the apparent 
singularity at $t=0$. It is understood that before doing so one has 
to express $\phi(t)$ in the coordinate $t$.
\be
\phi(t) = \phi(z = t^{-1}) \, \frac{1}{t^4} \,.
\ee

\subsection{The derivatives}
\label{derivatives_s5}

In order to compute amplitudes involving not only tachyons, we 
need the derivatives $\frac{\partial a_i}{\partial \xi_j}$ and 
$\frac{\partial a_i}{\partial \bar{\xi_j}}$. We don't have expressions 
for these quantities that we could directly evaluate, so we need to 
estimate these derivatives. We are using a Richardson's formula of 
order four, which for a function $f$ differentiable at least five times, 
reads
\be
f'(x) = \frac{1}{6h} \left(f(x-h) - 8 f(x-h/2) + 8 f(x+h/2) - f(x+h) \right)
+ {\cal O}(h^4) \,.
\label{Richardson} \ee
As we are working with numbers in double precision (which is usually around 
$16$ significant digits) we see from this formula that the most efficient $h$ 
is approximately $h = 10^{-3}$, which should give a result accurate to 
about twelve significant digits; taking a smaller $h$ would reduce 
the precision because of cancellation errors.
If we write $\xi_i = x_i + i \, y_i$, we have $\partial_{\xi_i} = 
\frac{1}{2} (\partial_{x_i} - i \, \partial_{y_i})$ and $\partial_{\bar{\xi_i}} = 
\frac{1}{2} (\partial_{x_i} + i \, \partial_{y_i})$, for each $a_i$ we thus 
need to evaluate the four derivatives $\frac{\partial a_i}{\partial {x_j}}$ and 
$\frac{\partial a_i}{\partial {y_j}}$, $j=1,2$. Each derivative calculated with 
(\ref{Richardson}) requires four evaluation of the function, so in total we 
need to solve the Strebel differential at sixteen points in order to evaluate 
accurately its derivatives at a given point.
Note that for the quartic vertex \cite{quartic}, the derivatives are 
calculated (\cite{Yang:2005ep, Yang:2005rx, Moe-Yang}) by 
differentiating the fit $a^{\rm fit}(\xi, \bar{\xi})$, but 
for the quintic we haven't been able to write a reasonably short fit.

\sectiono{The reduced moduli space}
\label{ms_s5}

In this section we describe the boundary of the reduced moduli space
${\cal V}_{0,5}$ (which we sometimes call simply moduli space, without
ambiguity since we always consider the reduced moduli space), and the
methods to evaluate it numerically. To represent this four-dimensional
moduli space, we first study its projection on the $\xi_1$-plane, this
can be done algebraically. Then at every point of the projection, we
have to describe the two-dimensional section in the $\xi_2$-plane; we
do this numerically.

\paragraph{}
We naturally want to reduce the numerical part of the problem to the
minimum. For this, we need to figure out what is the smallest part of
${\cal V}_{0,5}$ that we need to describe numerically and still be
able to integrate over the whole ${\cal V}_{0,5}$. We can do two
successive partitioning of the moduli space. Firstly, let five string
states $|\Psi_i\rangle$, $i=1,\ldots,5$, scattering on a
five-punctured sphere, given by a particular point $m$ of ${\cal
V}_{0,5}$. We can draw, on the sphere, the critical graph of the
Jenkins-Strebel quadratic differential uniquely determined by $m$. For
almost all $m$ (i.e. all $m \in {\cal V}_{0,5}$ except for the ones
belonging to a subset of measure zero), the graph will delimit a prism
consisting of two opposing triangles and three quadrilaterals (see
Figure \ref{prism_f5}); we can thus partition the moduli space into
ten regions, according to which three of the five states belong to the
quadrilateral faces of the prism. We conformally map
the sphere onto a sphere, requiring that the three quadrilateral
vertices are mapped to the standard points $z=0$, $z=1$ and $z=\infty$
(there are six such maps, but which one we choose is irrelevant as
will become clear below). 
Secondly, as in the case of the quartic vertex
\cite{Belo, quartic}, we can use the six ${\rm PSL}(2, {\mathbb C})$
transformations that permute the three points $\{0,1,\infty\}$, and
complex conjugation, to partition further the ten regions of ${\cal
V}_{0,5}$ into twelve parts. In total, the moduli space is thus
composed of $120$ parts, and we need to describe only one of them,
which we denote ${\cal A}_5$. The integration over the whole reduced
moduli space can then be written
\ben
&& \int_{{\cal V}_{0,5}} d\lambda_1 \wedge \ldots \wedge d\lambda_4 \langle \Sigma| 
b(v_{\lambda_1}) \ldots b(v_{\lambda_4}) |\Psi_1\rangle |\Psi_2\rangle 
|\Psi_3\rangle |\Psi_4\rangle |\Psi_5\rangle  = 
\label{intV05} \\
&& \hspace{-24pt} = \left( \int_{{\cal A}_5} + \int_\frac{1}{{\cal A}_5} + 
\int_{1-{\cal A}_5} + \int_\frac{1}{1-{\cal A}_5} + 
\int_{1-\frac{1}{{\cal A}_5}} + \int_\frac{{\cal A}_5}{{\cal A}_5-1} \right)
\Bigl( F(\Psi_1, \Psi_2, \Psi_3 | \Psi_4, \Psi_5) + {\rm permutations} \Bigr)
+ {\rm c.c.} \,,
\nonumber
\een
where $\lambda_i$ are real coordinates of the reduced moduli space
${\cal V}_{0,5}$, and the surface state $\langle \Sigma|$ corresponds
to the $5$-punctured sphere given by the parameters $\lambda_i$. The
antighost insertions $b(v_{\lambda_i})$ are not needed in this
section, they will be given in Section \ref{s_d5contact}. We denote
\be
F(\Psi_1, \Psi_2, \Psi_3 | \Psi_4, \Psi_5) \equiv 
d\lambda_1 \wedge \ldots \wedge d\lambda_4 \langle \Sigma| 
b(v_{\lambda_1}) \ldots b(v_{\lambda_4}) |\Psi_1\rangle |\Psi_2\rangle 
|\Psi_4\rangle |\Psi_5\rangle |\Psi_3\rangle \,,
\ee
and it is understood that the first three states are inserted on the
quadrilateral faces (i.e. the punctures $z_1=0$, $z_2=1$ and
$z_5=\infty$). The integrand on the second line of (\ref{intV05}) is
the sum of the ten permutations of $F(\Psi_1, \Psi_2, \Psi_3 | \Psi_4,
\Psi_5)$, each permutation being determined by which states are
assigned to the first three arguments (i.e. the quadrilaterals faces).
We are then left with six integrals that can all be written as 
integrals over ${\cal A}_5$ after pulling back their integrands through 
the aforementioned ${\rm PSL}(2, {\mathbb C})$ maps, which are explicitly 
\be 
z \rightarrow z \ , \quad z \rightarrow \frac{1}{z} \ , \quad 
z \rightarrow 1-z \ , \quad z \rightarrow \frac{1}{1-z} \ , \quad 
z \rightarrow 1 - \frac{1}{z} \ , \quad z \rightarrow \frac{z}{z-1} \,.
\label {PSL2C} \ee
The six other integrals, over the complex conjugates of the six domains, 
can be trivially pulled back to integrals over the six domains because the 
transformation is simply complex conjugation. This contribution 
is therefore the complex conjugate in (\ref{intV05}).

\begin{figure}[!ht]
\begin{center}
\input{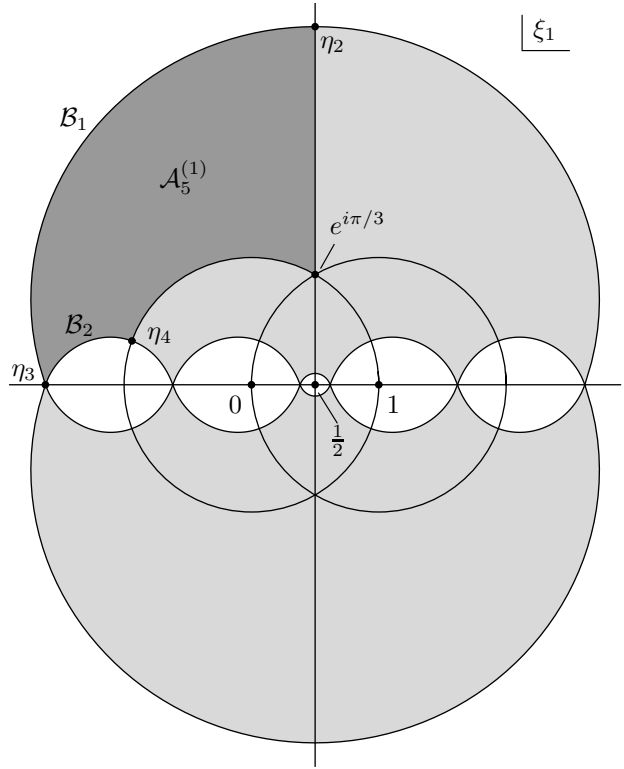}
\caption{\footnotesize{The projection of ${\cal V}_{0,5}^{\{0,1,\infty\}}$ 
on the $\xi_1$-plane. The darker region is the projection of ${\cal A}_5$.}}
\label{M5_f5}
\end{center}
\end{figure}

\paragraph{}
Now we need to describe ${\cal A}_5$. Let us define ${\cal V}_{0,5}^{\{0,1,\infty\}}$ 
to be the part of ${\cal V}_{0,5}$ whose three quadrilaterals correspond to the 
punctures $z=0$, $z=1$ and $z=\infty$. Its projection on the $\xi_1$-plane is shown 
on Figure \ref{M5_f5}.
The six mappings (\ref{PSL2C}), and complex conjugation, transform
simultaneously $\xi_1$ and $\xi_2$, we could thus choose to fix either
one in a given region. We choose to keep $\xi_1$ in the region ${\cal
A}_5^{(1)}$, characterized by being above the real axis, on the left
of the line $\Re \xi_1 = \frac{1}{2}$ and outside the unit circle
centered on zero (see Figure \ref{M5_f5}); and $\xi_2$ is unconstrained by 
these mappings. So ${\cal A}_5^{(1)}$ is the projection of ${\cal A}_5$ 
on the $\xi_1$-plane. 

\begin{figure}[!ht]
\begin{center}
\input{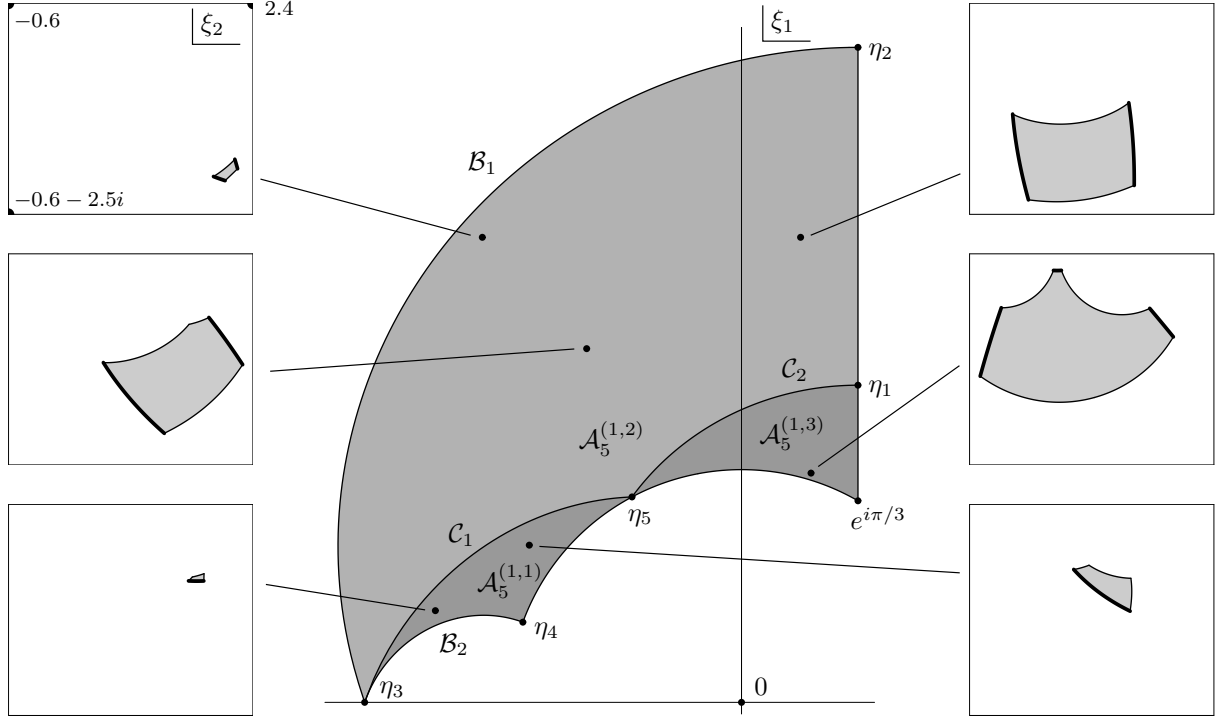}
\caption{\footnotesize{The projection of ${\cal A}_5$ on the 
$\xi_1$-plane. At various values of $\xi_1$, we show the section of the moduli 
space in the $\xi_2$-plane. The boxes all have the same scale. 
It is clearly visible that when we approach the boundaries ${\cal B}_1$ or ${\cal B}_2$, 
the section shrinks to zero. The thick boundaries of the sections are boundaries of 
the second kind, the others are of the first kind.}}
\label{ms_f5}
\end{center}
\end{figure}

\paragraph{}
We will now describe the remaining boundaries of ${\cal A}_5^{(1)}$, namely 
the two curves ${\cal B}_1$ and ${\cal B}_2$. In order to do this we must anticipate a 
little bit on the computation of the 
two-dimensional sections of ${\cal A}_5$ (the regions in the $\xi_2$-plane with 
fixed $\xi_1$). 
To make things clear, if we parameterize the four-dimensional moduli space by the 
four real coordinates $\left(\Re \xi_1, \Im \xi_1, \Re \xi_2, \Im \xi_2 \right)$, 
the section $S_{\xi_1}$ is the subset of the $\xi_2$-plane for which 
$\left(\Re \xi_1, \Im \xi_1, \Re \xi_2, \Im \xi_2 \right)$ is inside the 
moduli space.
We present now on Figure \ref{ms_f5}, 
a picture of ${\cal A}_5$, i.e. at various points $\xi_1$ of ${\cal
A}_5^{(1)}$ we show the section $S_{\xi_1}$. The boundaries of these
sections are of two kinds, the first kind is when an edge of the prism
has length $\pi$, while the boundaries of the second kind are found
when an edge has vanishing length (these are drawn with a thick line
on Figure \ref{ms_f5}). It is clear that when $\xi_1$ approaches the
boundaries ${\cal B}_1$ or ${\cal B}_2$, the section $S_{\xi_1}$ must
shrink to a point. As we will see below, the sections near ${\cal
B}_1$ have two boundaries of the second kind, so when this is reduced
to a point, we must have simultaneously two edges with length
zero. Near ${\cal B}_2$, the sections have only one boundary of the
second kind, but three boundaries of the first kind. So on ${\cal
B}_2$ we must have simultaneously one length zero and three lengths
$\pi$. As can be checked on Figure \ref{prism_f5} and
Equ.(\ref{lengths-conditions}) by exhaustion of all possibilities,
this implies that two edges have vanishing lengths. In other words,
{\em a sphere on the boundary ${\cal B}_1$ or ${\cal B}_2$ is
described by a quadratic differential with two double zeros}. We now
understand the importance of the quadratic differentials discussed in
Section \ref{twodouble_s5}, and we are glad that they could be solved
in terms of the algebraic curve $T(\ell)$ Equ.(\ref{Tell}).

\paragraph{}
What we must do now is to list all the conformal and anti-conformal
maps $z = h(w)$ such that the poles $\{-1,1,\alpha,-\alpha,\infty\}$
of the quadratic differentials with two double zeros of Section
\ref{twodouble_s5} are mapped to the poles in the $z$-plane
$\{0,1,\xi_1,\xi_2,\infty\}$, and such that the critical graph (Figure
\ref{twodouble_f5}) is mapped to a critical graph that is obtainable
continuously from the graph of the right of Figure \ref{mostsym_f5} by
the vanishing of two lengths, and such that $\xi_1$ belongs to the
region characterized by having positive imaginary part, real part
smaller that $1/2$ and absolute value greater than one.  There are
four maps obeying all these requirements. The first one is
\be
h_1(w) = \overline{\left(\frac{\alpha-w}{2 \alpha}\right)} \,,
\ee
which maps 
\be
h_1(\infty) = \infty \ , \quad h_1(-\alpha) = 1 \ , \quad h_1(\alpha) = 0 \,, 
\ee
and $\xi_1$ and $\xi_2$ must therefore be the images of $-1$ and $1$.
\be
\xi_1 = h_1(-1) = \overline{\left(\frac{\alpha+1}{2 \alpha}\right)} 
\ , \quad \xi_2 = h_1(1) = 
\overline{\left(\frac{\alpha-1}{2 \alpha}\right)} = 1 - \xi_1 \,.
\ee
This corresponds to the curve ${\cal B}_1$
\be
\boxed{{\cal B}_1 = \left\{ \overline{\left(
\frac{\alpha(\ell)+1}{2 \alpha(\ell)} \right)} \ , \ \ell \in 
\left[ 0, \frac{\pi}{2} \right] \right\}} \,,
\ee
where $\alpha(\ell)$ is given by (\ref{alphal}). The second map is 
\be
h_2(w) = \frac{\alpha-1}{\alpha+1} \frac{w-1}{w+1} \,.
\ee
It maps 
\ben
&& h_2(1) = 0 \ , \quad h_2(-1) = \infty \ , \quad h_2(-\alpha) = 1 \,,
\nonumber \\
&& \xi_1 = h_2(\infty) = \frac{\alpha-1}{\alpha+1} \ , \quad
\xi_2 = h_2(\alpha) = \xi_1^2 \,.
\een
This defines the curve ${\cal B}_2$
\be
\boxed{{\cal B}_2 = \left\{ 
\frac{\alpha(\ell)-1}{\alpha(\ell) +1} \ , \ \ell \in 
\left[ 0, \frac{\pi}{2} \right] \right\}} \,.
\ee
The third map is
\be
h_3(w) = \frac{2}{\alpha+1} \frac{w-\alpha}{w-1} \,,
\ee
for which
\ben
&& h_3(1) = \infty \ , \quad h_3(-1) = 1 \ , \quad h_3(\alpha) = 0
\nonumber \\
&& \xi_1 = h_3(-\alpha) = \frac{4 \alpha}{(\alpha+1)^2} 
\ , \quad \xi_2 = h_3(\infty) = \frac{2}{\alpha+1} \,.
\een
We will call the corresponding curve ${\cal C}$
\be
{\cal C} = \left\{ 
\frac{4 \alpha(\ell)}{(\alpha(\ell) + 1)^2} \ , \ \ell \in 
\left[ 0, \frac{\pi}{2} \right] \right\} \,.
\ee
Before we discuss the fourth map and the meaning of the third one, we 
need to look at what we have found so far. At the special point $\ell = 0$, we 
recall (\ref{alphabounds}) that $\alpha(0) = 2 - \sqrt{5}$ and thus we find 
that all three curves 
meet at the point $\eta_3$ (Figure \ref{ms_f5}), where
\be
\eta_3 = \left. {\cal B}_1 \right|_{\ell=0} = \left. {\cal B}_2 \right|_{\ell=0} = 
\left. {\cal C} \right|_{\ell=0} = -\frac{1+\sqrt{5}}{2} \,.
\ee
When $\ell = \frac{\pi}{2}$, the curve ${\cal B}_1$ meets the axis 
$\Re \xi_1 = \frac{1}{2}$ at the point $\eta_2$. From (\ref{alphabounds}) we find
\be
\eta_2 = \frac{1}{2} + 
\frac{32 \sqrt{6} + 19 \sqrt{15}}{54} \, i \,.
\ee
Also when $\ell = \frac{\pi}{2}$, the curve ${\cal B}_2$ meets the unit 
circle at the point $\eta_4$, given by
\be
\eta_4 = 
\frac{27 + \left( 32 \sqrt{6} + 19 \sqrt{15} \right) \, i}
{27 - \left( 32 \sqrt{6} + 19 \sqrt{15} \right) \, i} \,.
\ee
The situation for the curve ${\cal C}$ is a little bit different because for 
$\ell > \ell_0 \approx 0.67622$, the curve is outside of the allowed domain 
because it is inside the unit circle centered at the origin. But we can map the 
piece of curve with $\ell > \ell_0$ back inside the domain by the map 
$z \rightarrow 1/\bar{z}$. So the curve ${\cal C}$ gives us actually two 
curves ${\cal C}_1$ and ${\cal C}_2$ (and the fourth map is just 
$h_4(w) = 1/\overline{h_3(w)}$). We have thus
\be
{\cal C}_1 = \left\{ 
\frac{4 \alpha(\ell)}{(\alpha(\ell) + 1)^2} \ , \ \ell \in 
\left[ 0, \ell_0 \right] \right\} \quad , \qquad 
{\cal C}_2 = \left\{ 
\overline{\left( \frac{(\alpha(\ell) + 1)^2}{4 \alpha(\ell)} \right)} \ , \ \ell \in 
\left[ \ell_0, \frac{\pi}{2} \right] \right\} \,.
\ee
At $\ell = \ell_0$ these two curves intersect on the unit circle at the 
point 
\be
\eta_5 \approx -0.47019 + 0.88256 \, i \,.
\ee
At last, when $\ell = \frac{\pi}{2}$, the curve ${\cal C}_2$ meets the axis 
$\Re \xi_1 = \frac{1}{2}$ at the point $\eta_1$, with
\be
\eta_1 = \frac{1}{2} + \frac{19 \sqrt{15}}{54} \, i \,.
\ee
\begin{figure}[!ht]
\begin{center}
\input{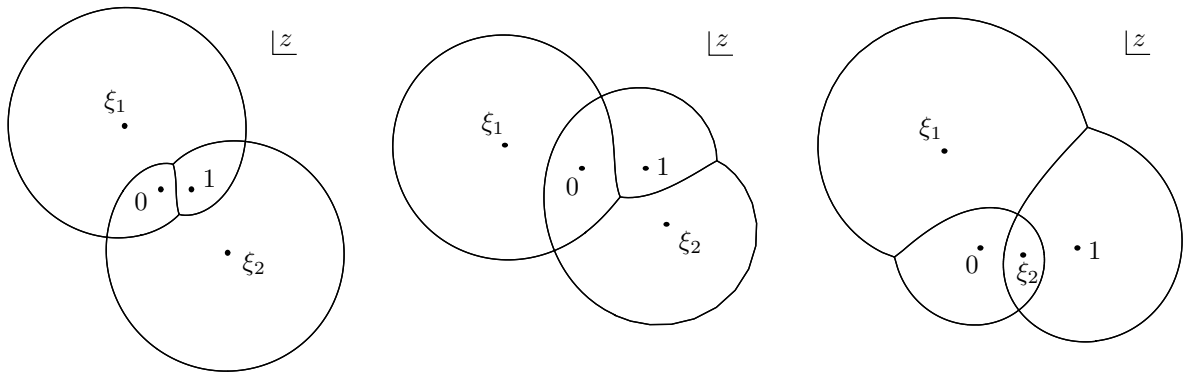}
\caption{\footnotesize{The critical graphs of the quadratic 
differentials obtained by mapping the quadratic differential of Figure 
\ref{twodouble_f5} with, respectively, $h_1$, $h_2$ and $h_3$.}}
\label{B1B2C2_f5}
\end{center}
\end{figure}
In Figure \ref{B1B2C2_f5} we show the critical graphs of the quadratic
differentials obtained by mapping the quadratic differential of Figure
\ref{twodouble_f5} with, respectively, $h_1$, $h_2$ and $h_3$. So they
are respectively on ${\cal B}_1$, ${\cal B}_2$, and ${\cal C}_2$.

\paragraph{} 
It remains now to understand the meaning of the curves ${\cal C}_1$
and ${\cal C}_2$. From Figure \ref{B1B2C2_f5} we can be more specific
by noting that on these curves, one quadrilateral and one triangle
edge are zero. By exhausting all combinations on the prism of Figure
\ref{prism_f5} and Equ.(\ref{lengths-conditions}), we see that this
implies that we have also three triangle edges of length $\pi$. We
thus have an intersection of three boundaries of the first kind and
one boundary of the second kind.  This can happen if a boundary of the
second kind of the section $S_{\xi_1}$ is reduced to a
point. Therefore the curves ${\cal C}_1$ and ${\cal C}_2$ partition
${\cal A}_5^{(1)}$ into three regions ${\cal A}_5^{(1,n)} \, , \
n=1,2,3$ in which the sections $S_{\xi_1}$ have exactly $n$ boundaries
of the second kind. This partition is shown on Figure \ref{ms_f5},
where we can also verify the number of boundaries of the sections
shown in the boxes. We do not prove that this partitioning is as
shown, but we have checked it numerically beyond doubt.

\paragraph{}
To end this section, we briefly sketch how we plot numerically the
sections. We start by plotting the boundaries of the second kind; this
is done with the quadratic differentials with one double zero, studied
in Section \ref{onedouble_s5}. The end of the boundaries are detected
when a length is $\pi$, and accurately determined. We then start from
one end (where, say, $\ell_i =\pi$) to plot a boundary of the first
kind by finding the points $\xi_2$ for which $\ell_i = \pi$. We end
either when another length (say $\ell_j$) becomes $\pi$, or when we
end up very close to a boundary of the second kind. In the first case,
we determine accurately the corner (the intersection of two
boundaries) and continue by looking at the points for which
$\ell_j=\pi$.  In the second case we go on by plotting the boundary of
the first kind attached to the other end of the boundary of the second
kind that we just met.  We continue this process until we arrive at
the other end of the second-kind boundary from which we started. The
result will be a closed curve given by a set of points separated by an
approximately fixed distance $h$; and the special points at the
corners of the curve are always precisely given.

\sectiono{The five-tachyon contact term}
\label{s_t5}

To calculate the five-tachyon contact term, we take the general
formula established in \cite{Belo-Zwie} for the $N$-tachyon term (or
use Equ.(\ref{N-amplitude})). Namely

\be
\kappa^2 V_{t^N} = \frac{(-1)^{N-1}}{N!} \frac{2}{\pi^{N-3}} \int_{{\cal V}_{0,N}} 
\prod_{i=1}^{N-3} 
\frac{dx_i \wedge dy_i}{\rho_i^2} \frac{1}{\rho_{N-2}^2(0) \rho_{N-1}^2(1) \rho_N^2(\infty)}
\ee
where $\xi_i = x_i + i \, y_i$.
Since the five external states are the same, the integration over ${\cal V}_{0,5}$ can 
be written as $120$ times the integration over ${\cal A}_5$. Therefore
\be
\kappa^2 V_{t^5} = \frac{2}{\pi^2} \int_{{\cal A}_5} 
dx_1 \wedge dy_1 \wedge dx_2 \wedge dy_2 \, \mu(\xi_1, \bar{\xi_1}, \xi_2, \bar{\xi_2}) 
\quad , \qquad 
\mu \equiv 
\frac{1}{(\rho_1 \rho_2 \rho_3 \rho_4 \rho_5)^2} \,,
\ee

\paragraph{}
It is now a good place to explain how we perform the numerical
integration.  There are five distinct steps in the process. 

\paragraph{Step 1}
First we compute the boundary of the reduced moduli space. For this we
draw a covering rectangular grid of $(N+1) \times (N+1)$ points on
${\cal A}_5^{(1)}$, the projection of ${\cal A}_5$ on the
$\xi_1$-plane. Then at each of the points $\xi_1$ of the grid we {\em
attempt} to plot, in the $\xi_2$-plane, the boundary of the section
$S_{\xi_1}$. In the $\xi_2$-plane, this boundary is represented as a
list of points, and the space between two successive points is taken
to be approximately $h$. It may happen that the algorithm fails to
find a Strebel differential because the Newton method fails to
converge (for example if we are close to a singular quadratic
differential, near the boundaries ${\cal B}_1$ or ${\cal B}_2$). In
that case, the algorithm may be able to fix the problem by trying
other initial values; if this still fails the consequence may be that
the algorithm is unable to draw the section, but this is not dramatic,
it simply keeps record of this failure and proceeds to the next point
of the grid. We are using three different grids which have
respectively $(N,h) = (30,0.1)$, $(70, 0.05)$ and $(100, 0.03)$.
\begin{figure}[!ht]
\begin{center}
\input{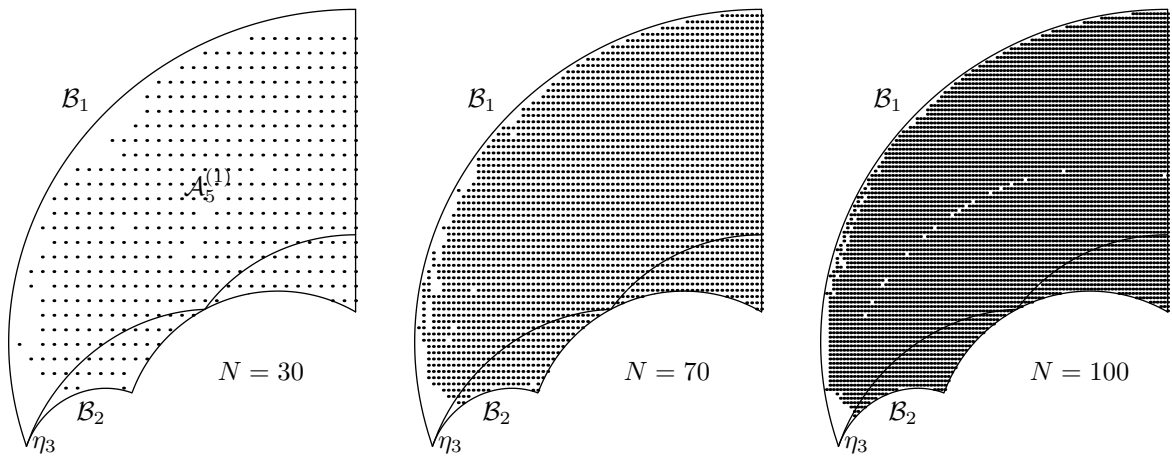}
\caption{\footnotesize{The projection of ${\cal A}_5$
covered with grids of $(N+1) \times (N+1)$ points with $N=30$, $70$
and $100$. The dots represent the points of the grids for which the
section could be computed.}}
\label{grids_f5}
\end{center}
\end{figure}
On Figure \ref{grids_f5} we show all the points of these three grids
for which the section could be plotted. We see that we have some gaps
near the boundary ${\cal B}_1$, especially near the point $\eta_3$
where it meets the boundary ${\cal B}_2$. This is not surprising
because at this point the quadratic differential has two triple zeros,
a very singular point indeed. But it is clear that as we increase the
number of points on the grid, the gaps are becoming smaller. This is
due to the fact that, going from one point to the next, the quadratic
differential solver based on the Newton method, can use the solutions
of the nearby points as seeds, and of course if they are closer the
Newton method will have more chance to converge.  The grid with
$N=100$ is however already at the limit of the computational power of
a desktop computer. We notice also a few gaps away from the
boundaries, and most noticeably there seems to be a line of gaps in
the middle of the grid $N=100$. The failure there is actually not due
to the Newton method but to the difficulty of plotting the boundary of
a section when one very small boundary of the first kind is stuck
between two other boundaries of the first kind. This could be fixed by
improving the algorithm, but we don't really need to do this at this
point because we will eventually use quadratic interpolation to fix
these gaps.  One may worry about the size of the gaps near ${\cal
B}_1$ and especially near $\xi_1 = \eta_3$, but we recall that near
the boundaries ${\cal B}_1$ and ${\cal B}_2$ the sections become
small, and therefore the gaps actually represent a small
four-dimensional volume, and will be fixed by extrapolation. The error
made by the interpolations and extrapolation will be estimated later.

\paragraph{Step 2} 
Now we cover every section with a rectangular grid, with spacing $s$
along the real and imaginary directions, and at every of these points
we attempt to find the Strebel quadratic differential. Again it is
possible that some differentials are not found, but this will be fixed
by interpolation (extrapolation). At every of these points, we store
all the data of the quadratic differential, i.e. $\xi_i$, $a_i$,
$\partial a_i / \partial \xi_j$, $\partial a_i / \partial
\bar{\xi_j}$, $i,j = 1,2$ and $\rho_I$, $I=1,\ldots,5$. For our three
particular grids $N=30$, $70$ and $100$, the spacing $s$ is chosen as
respectively $s=0.1$, $0.05$ and $0.04$. We also consider the same
grid $N=70$ with the smaller spacing $s=0.04$.

\paragraph{Step 3} 
We integrate $\mu$ on every sections. 
\be
M(\xi_1, \bar{\xi_1}) \equiv \int_{S_{\xi_1}} dx_2 \wedge dy_2 \, 
\mu(\xi_1, \bar{\xi_1}, \xi_2, \bar{\xi_2}) \quad , \quad 
\xi_2 = x_2 + i \, y_2 \,.
\ee
because the shapes of the sections 
are complicated, we use a Monte-Carlo technique and integrate the function
\be
\mu'(\xi_1, \bar{\xi_1}, \xi_2, \bar{\xi_2}) = 
\left\{ \begin{array}{l} \mu(\xi_1, \bar{\xi_1}, \xi_2, \bar{\xi_2}) \quad 
{\rm if} \quad \xi_2 \in S_{\xi_1} \\ 0 \quad {\rm otherwise} \end{array} \right.
\ee
over a rectangle containing the section $S_{\xi_1}$. To evaluate $\mu$
at random points $\xi_2$ we use quadratic interpolation over the
closest $3 \times 3$ sub-array of points of the grid. Actually, we
find it more natural to interpolate $\log \mu$, and exponentiate
the result of the interpolation. 

\paragraph{Step 4}
We can finally perform the whole integration
\be
\int_{{\cal A}_5} dx_1 \wedge dy_1 \wedge dx_2 \wedge dy_2 \, \mu(\xi_1, \bar{\xi_1}, 
\xi_2, \bar{\xi_2}) = 
\int_{{\cal A}_5^{(1)}} dx_1 \wedge dy_1 \, M(\xi_1, \bar{\xi_1}) \,.
\ee
We use again a Monte-Carlo technique, integrating
\be
M'(\xi_1, \bar{\xi_1}) = \left\{ \begin{array}{l} M(\xi_1, \bar{\xi_1}) \quad 
{\rm if} \quad \xi_1 \in {\cal A}_5^{(1)} \\ 0 \quad {\rm otherwise} \end{array} \right. \,,
\ee 
and integrate $M'$ over a rectangle containing ${\cal A}_5^{(1)}$. To
evaluate $M$, we use again quadratic interpolation of $\log M$, and
exponentiate the result of the interpolation; we observe that it is a
better interpolation that if we had simply interpolated $M$ itself.

\paragraph{Step 5: Error estimation}
We would like to calculate an estimate of the numerical uncertainty on the 
result. There are four sources of error, namely
\begin{enumerate}
\item The error due to the finite spacing $h$. In order to decide if
$\xi_2 \in S_{\xi_1}$, we check if $\xi_2$ is inside the polygon
formed by all points describing the boundary of $S_{\xi_1}$. This is
basically a linear interpolation, so the error made will be of the
order of $h^2$, a potentially large error. To reduce it, we are
computing the integral twice, once in the normal way, and a second
time with all the sections replaced with cruder ones with spacing $2
h$. Practically we are removing half the points of every sections but
keep the corners (the intersections between boundaries). In order to
isolate this source of errors from the other sources, it is important
to use the same Monte-Carlo samples in both integrations (i.e. the
sequence of random points must be the same). Let us denote $\kappa^2
V_{t^5}(h)$ and $\kappa^2 V_{t^5}(2h)$ the results of these two
integrations, and let $\kappa^2 V_{t^5}$ be the exact value. Since the 
error is quadratic in $h$, we have 
\be
\kappa^2 V_{t^5}(2h) - \kappa^2 V_{t^5} = 4 \left( 
\kappa^2 V_{t^5}(h) - \kappa^2 V_{t^5} \right) + {\cal O}(h^3) \,,
\ee
or in other words 
\be
\kappa^2 V_{t^5} = \frac{1}{3} \left( 4 \kappa^2 V_{t^5}(h) -
\kappa^2 V_{t^5}(2h) \right) + {\cal O}(h^3) \,,
\ee
and it will turn out that we can neglect the residual error as it will be smaller 
than the other sources.

\item The error done by the interpolation and extrapolation of $\mu$. Let us
write $\mu = \mu_{\rm inter} + \delta_{\mu}$. We will then treat this
source of error as systematic, in other words
\be
\sigma_2 \equiv \int_{{\cal A}_5} dx_1 \wedge dy_1 \wedge dx_2 \wedge dy_2 \, 
\left|\delta_{\mu}(\xi_1, \bar{\xi_1}, \xi_2, \bar{\xi_2})\right| \,.
\ee
The quadratic interpolation is done with the routine {\tt polin2} of 
\cite{NR}. Based on Neville's algorithm, it can give an 
estimate of $\delta_{\mu}(\xi_1, \bar{\xi_1}, \xi_2, \bar{\xi_2})$.

\item the error due to the interpolation and extrapolation of $M$. It is
a systematic error as well. We write $M = M_{\rm inter} + \delta_M$ and 
\be
\sigma_3 \equiv \int_{{\cal A}_5^{(1)}} dx_1 \wedge dy_1 \, 
\left| \delta_M \right|
\ee

\item The error coming from the Monte-Carlo integrations. Those are
purely statistical errors, and the errors coming from the integrations
in Step 3 will cancel down to a negligible quantity in the final
result if they are done with enough samples. We are thus only
considering the Monte-Carlo error $\sigma_{MC}$ on the last integral
of Step 4.
\end{enumerate}
So finally
\be
\sigma^2_{\kappa^2 V_{t^5}} = \frac{2}{\pi^2} \left( \sigma_2^2 + \sigma_3^2 + 
\sigma_{MC}^2 \right) \,,
\ee
and we will take enough samples in the Monte-Carlo integration so that
the error will be dominated by the errors on interpolation and
extrapolation, which depend on the finesse of the grid.

\paragraph{}
We note that Steps 1 and 2 need to be done only once for every grid. Once these 
have been done, every integration requires only Steps 3, 4 and 5.

\paragraph{} 
We now state our results. We use four different grids, with $(N,h,s)$
respectively equal to $(30,0.1,0.1)$, $(70,0.05,0.05)$,
$(70,0.05,0.04)$ and $(100,0.03,0.04)$. These last two grids are at
the computational limit of a desktop computer with our C++ code.
Indeed it takes several days to compute $a_i(\xi_1, \bar{\xi_1},\xi_2,
\bar{\xi_2})$ and their derivatives and $\rho_I(\xi_1,
\bar{\xi_1},\xi_2, \bar{\xi_2})$ at every points of the
four-dimensional grid $(70,0.05,0.04)$ (which has approximately 1.6
million points inside ${\cal A}_5$, and fills 1.3 GBytes of RAM). The
grid $(100,0.03,0.04)$ is actually too large to store the derivatives
of $a_i$, so we can use it only for the computation of $\kappa^2
V_{t^5}$. The computation of the five-dilaton contact term in Section
\ref{s_d5} will have to be done with the grid $(70,0.05,0.04)$. All
the Monte-Carlo integrations are done with one iteration of the
routine {\tt vegas} of \cite{NR}. The integrations of Step 3 are done
with $10^5$ samples, and the integrations of Step 4 are done with
$10^6$ samples. Our results are shown in Table \ref{t5-table}.
\begin{table}[!ht]
\begin{center}
\begin{tabular}{|c||c|c|c|c|}
\hline
$(N,h,s)$ & $(30,0.1,0.1)$ & $(70,0.05,0.05)$ & $(70,0.05,0.04)$ & $(100,0.03,0.04)$ \\
\hline \hline
$\kappa^2 V_{t^5}(2h)$ & $10.0774 \pm 0.065$ & $9.95537 \pm 0.013$ & $9.95197 \pm 0.010$ & 
$9.93412 \pm 0.008$ \\
\hline
$\kappa^2 V_{t^5}(h)$ & $10.0045 \pm 0.065$ & $9.93535 \pm 0.013$ & $9.93252 \pm 0.010$ & 
$9.92648 \pm 0.008$ \\
\hline \hline
$\kappa^2 V_{t^5}$ & $9.980 \pm 0.065$ & $9.929 \pm 0.013$ & $9.926 \pm 0.010$ & 
$9.924 \pm 0.008$ \\
\hline
\end{tabular}
\caption{\footnotesize{The results of the integrations and their uncertainties on 
our three different grids.}}
\label{t5-table}
\end{center}
\end{table}
It is reassuring that all four results are compatible within their
error bounds. They are not statistically independent however, and we
should therefore not combine them. So we take the result from the
finest grid as final answer
\be
\boxed{\kappa^2 V_{t^5} = 9.924 \pm 0.008} \,.
\label{t5} \ee

\sectiono{The five-dilaton effective term}
\label{s_d5}

The goal of this section is to check the validity of our
computation. The quartic vertex computed in \cite{quartic} was checked
in \cite{Yang:2005iu} with marginal fields. A similar analysis is
unfortunately not possible here because we can couple only an even
number of marginal fields like $\alpha_{-1} \bar{\alpha}_{-1} c_1
\bar{c}_1 |0\rangle$ and the vanishing of the quintic term of its
effective potential is trivial. However we can compute the amplitude
of five dilatons $|D\rangle$, where
\be
|D\rangle = \left(c_1c_{-1}-\bar{c}_1 \bar{c}_{-1}\right) |0\rangle \,,
\label{dilaton}
\ee
and compare it to the prediction of the dilaton theorem. 

\begin{figure}[!ht]
\begin{center}
\input{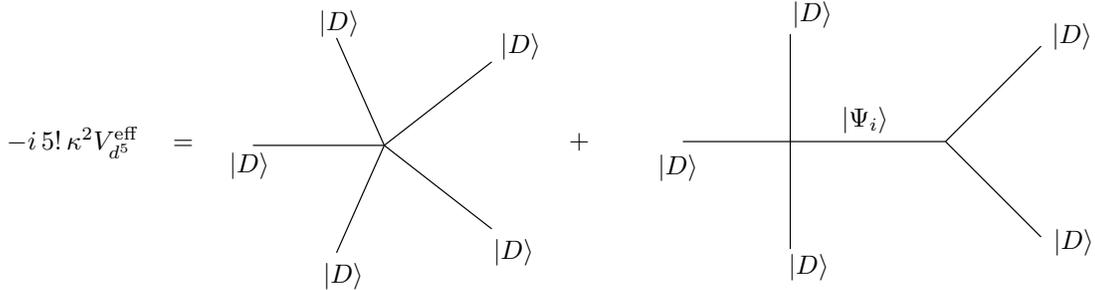}
\caption{\footnotesize{The term $d^5$ of the dilaton effective potential. The fields 
$|\Psi_i\rangle$ that propagate in the second diagram, are the tachyon and 
all the massive fields, but not the dilaton.}}
\label{Feynman_f5}
\end{center}
\end{figure}

\paragraph{}
The dilaton theorem states that the effective potential of the dilaton
should be identically zero. That the $V^{\rm eff}_{d^3}$ term vanishes
is due to the fact that the cubic amplitude of three dilatons is zero
because the ghost numbers do not work out. The vanishing of $V^{\rm
eff}_{d^4}$ is nontrivial; it has been checked in \cite{Yang:2005ep}
that the contribution from the quartic amplitude $\kappa^2 V_{d^4}$
cancels the contribution from Feynman diagrams with two cubic
vertices. This cancellation was found to be good to about $0.2\%$,
furnishing a very good evidence that the quartic computations are done
right.  In this section, we want to check the vanishing of $V^{\rm
eff}_{d^5}$. There are two contributions to this effective term shown
on Figure \ref{Feynman_f5},
namely the contact term $\kappa^2 V_{d^5}$, and the Feynman term 
${\cal C}_{d^5}$
\be
\kappa^2 V^{\rm eff}_{d^5} = \kappa^2 V_{d^5} + {\cal C}_{d^5} \,.
\ee
We will start by evaluating the Feynman contribution, then we will compute the 
contact term with the machinery developed in this paper.

\subsection{The Feynman contribution}

The three-string vertex can couple only an even number of states with
asymmetric ghost numbers like the dilaton. So the diagram consisting
of three three-string vertices doesn't contribute, and the only
Feynman diagram contributing to ${\cal C}_{d^5}$ is therefore the one
shown on the right of Figure \ref{Feynman_f5}.  To get the right
relative sign between this diagram and the contact term, we must keep
track of all the coefficients. Since the vertices bring in the
amplitude a factor $-i$, and the propagator a factor $i$, we have
\ben
&& (-i) \, 5! \, {\cal C}_{d^5} = \binom{5}{3}
\sum_{i,j\neq 2} (-i)\left\{D,D,D,\Psi_i\right\} (i) \, \left(-M^{-1}\right)_{ij} 
(-i)\left\{ D,D,\Psi_j\right\} \,,
\nonumber \\
&& \Rightarrow {\cal C}_{d^5} = -\frac{1}{12}
\sum_{i,j\neq 2} \left\{D,D,D,\Psi_i\right\} \left(M^{-1}\right)_{ij} 
\left\{ D,D,\Psi_j\right\} \,,
\label{Cd5}
\een
where $(-M^{-1})$ is the zero-momentum propagator for all the fields
except the dilaton. $M$ is therefore given by the quadratic term
\be
M_{ij} = \langle \Psi_i | c_0^- Q_{B} |\Psi_j \rangle \ , \quad i,j\neq 2 \,. 
\ee 
Note that we work in a non-orthonormal
basis, and $M$ is therefore not diagonal. ${\cal C}_{d^5}$ must be evaluated 
with level truncation, we will do explicitly the first few levels. First it is 
convenient to decompose ${\cal C}_{d^5}$ into the sum of contributions at 
every level
\be
{\cal C}_{d^5}(\ell) = \sum_{\ell' \leq \ell} c_{d^5}(\ell') \,.
\ee
And we will write $M_\ell$ for the matrix $M$ with only fields of level $\ell$, 
and we define the row vectors
\ben
C_\ell &\equiv& \left( \left\{ D,D,\Psi_i \right\} \right) \nonumber \\
Q_\ell &\equiv& \left( \left\{ D,D,D,\Psi_i\right\} \right) \,,
\een
where the index $i$ runs over all fields of level $\ell$. With these notations, 
we can write
\be
c_{d^5}(\ell) = - \frac{1}{12} Q_\ell \cdot M_\ell^{-1} \cdot C_\ell^T \,.
\label{cd5}
\ee

\paragraph{}
At level zero, only the tachyon propagates, and we thus have
\be
{\cal C}_{d^5}(0) = - \frac{1}{12} \left\{D,D,D,T\right\} M_0^{-1} 
\left\{D,D,T \right\} = - \frac{1}{12} (-5.716) (-\frac{1}{2}) (-\frac{27}{16}) = 
0.4020 \,.
\ee

\paragraph{}
At level four, the four fields are (we are using the notations of \cite{Moe-Yang})
\ben
&& |\Psi_3\rangle = \left(b_{-2}c_1\bar{c}_{-2}\bar{c}_1 - 
\bar{b}_{-2}\bar{c}_1 c_{-2} c_1 \right) \vac \ , \quad
|\Psi_4\rangle = c_{-1} \bar{c}_{-1} \vac \nonumber \\
&& |\Psi_5\rangle = L_{-2}c_1 \bar{L}_{-2} \bar{c}_1 \vac \ , \quad
|\Psi_6\rangle = \left( c_{-1} \bar{L}_{-2} \bar{c}_1 - 
\bar{c}_{-1} L_{-2} c_1 \right) \vac \,.
\een
And we have
\ben
M_4 &=& {\rm diag}\left(-4,2,338,-52\right)
\nonumber \\
C_4 &=& \left( 0, -\frac{1}{48}, -\frac{4225}{432}, 
\frac{65}{72} \right)
\nonumber \\
Q_4 &=& \left(-4.547, -1.811, -22.33, 10.69 \right)
\een
The first two lines are done with standard techniques, and the third line has 
been calculated in \cite{Moe-Yang}, we refer the reader to \cite{Moe-Yang} and 
\cite{Yang:2005ep} for the 
details of this computation. With these values (\ref{cd5}) gives the contribution 
of level four 
\be
c_{d^5}(4) = -0.03996
\ee
which gives the total contribution
\be
{\cal C}_{d^5}(4) = {\cal C}_{d^5}(0) + c_{d^5}(4) = 0.3620 \,.
\ee

\paragraph{}
At level six, the fields are 
\ben
&&|\Psi_7\rangle = \left(b_{-2} \bar{c}_{-2} \bar{c}_{-1} \bar{c}_{1} -
\bar{b}_{-2} c_{-2} c_{-1} c_{1} \right) \vac \ , \quad
|\Psi_8\rangle = \left(L_{-2} \bar{c}_{-3} \bar{c}_1 - 
\bar{L}_{-2} c_{-3} c_{1} \right) \vac \nonumber \\
&&|\Psi_9\rangle = L_{-2} \bar{L}_{-2} \left(\bar{c}_{-1}\bar{c}_1  - 
c_{-1} c_1 \right) \vac \ , \quad
|\Psi_{10}\rangle = \left(b_{-3} c_{1} \bar{c}_{-3} \bar{c}_{1} -
\bar{b}_{-3} \bar{c}_{1} c_{-3} c_{1} \right) \vac \nonumber \\
&&|\Psi_{11}\rangle = \left( b_{-3} c_1 \bar{L}_{-2} \bar{c}_{-1} \bar{c}_{1} -
\bar{b}_{-3} \bar{c}_1 L_{-2} c_{-1} c_{1} \right) \vac \ , \quad 
|\Psi_{12}\rangle = c_{-2} \bar{c}_{-2} \vac \ , \quad 
|\Psi_{13}\rangle = L_{-3} c_1 \bar{L}_{-3} \bar{c}_1 \vac \nonumber \\
&&|\Psi_{14}\rangle = b_{-2} c_{-1} c_{1} \bar{b}_{-2} \bar{c}_{-1} \bar{c}_{1}
\vac \ , \quad |\Psi_{15}\rangle = \left( c_{-2}\bar{L}_{-3}\bar{c}_{1} - 
\bar{c}_{-2}L_{-3}c_{1} \right) \vac \nonumber \\
&&|\Psi_{16}\rangle = \left(c_{-2}\bar{b}_{-2}\bar{c}_{-1}\bar{c}_{1} - 
\bar{c}_{-2}b_{-2}c_{-1}c_{1} \right) \vac \ , \quad 
|\Psi_{17}\rangle = \left(L_{-3}c_{1}\bar{b}_{-2}\bar{c}_{-1}\bar{c}_{1}  - 
\bar{L}_{-3}\bar{c}_{1}b_{-2}c_{-1}c_1 \right) \vac \nonumber \,,
\een
and we have
\ben
M_6 &=& \left( \begin{array}{c} 8 \end{array} \right)
\oplus \left( \begin{array}{cccc} 
0 &     0 & 0 & -104 \\ 
0 & -1352 & 0 &    0 \\
0 &     0 & -8 & 0 \\
-104 & 0 & 0 & 0 \end{array} \right)
\oplus \left( \begin{array}{ccc}
0 & 0 & 4 \\
0 & 10816 & 0 \\
4 & 0 & 0 \end{array} \right)
\oplus \left( \begin{array}{ccc}
0 & 0 & 416 \\
0 & 8 & 0 \\
416 & 0 & 0 \end{array} \right)
\nonumber \\
C_6 &=& \frac{100}{729} \left(-8, 0,0,0,0,1,0,4,0,4,0 \right) \\
Q_6 &=& \left(-2.103, 2.363, -20.92, -0.4412, 7.090, 0.1251, -49.67, 
0.4998, 4.325, 1.603, -8.649\right) \,. \nonumber 
\een
And we find
\be
c_{d^5}(6) = -0.03606 \,,
\ee
which gives the total contribution
\be
{\cal C}_{d^5}(6) = {\cal C}_{d^5}(4) + c_{d^5}(6) = 0.3259 \,.
\ee

With the techniques developed in 
\cite{quartic,Yang:2005rx,Yang:2005ep,Yang:2005iu,Moe-Yang}, we can 
evaluate $\left\{D,D,D,\Psi_i\right\}$ for fields $\Psi_i$ of 
level up to ten. The results for ${\cal C}_{d^5}(\ell)$, $\ell = 0, 4, 6, 8, 10$ 
are shown in Table \ref{Cd5-table}. 
\begin{table}[!ht]
\begin{center}
\begin{tabular}{|c||c|c|c|c|c|}
\hline
$\ell$ & $0$ & $4$ & $6$ & $8$ & $10$ \\
\hline
${\cal C}_{d^5}(\ell)$ & $0.4020$ & $0.3620$ & $0.3259$ & $0.3167$ & $0.3112$ \\
\hline
\end{tabular}
\caption{\footnotesize{The value of ${\cal C}_{d^5}$ with propagating fields 
up to level $\ell$.}}
\label{Cd5-table}
\end{center}
\end{table}
We will follow the method used in \cite{Yang:2005ep} to extrapolate the above 
results to $\ell = \infty$. Namely we use a fit of the form
\be
{\cal C}_{d^5}(\ell) = f_0 + \frac{f_1}{\ell^\gamma} \,.
\label{fit} \ee
In fitting ${\cal C}_{d^4}(\ell)$, the authors of \cite{Yang:2005ep} found that 
$\gamma=3$ give the 
best results. Here by fitting the four values ${\cal C}_{d^5}(\ell)$, $\ell = 4,6,8,10$ with 
(\ref{fit}), we find that $\gamma = 2.51$. We take this as an evidence that the data should be 
fitted with 
\be
{\cal C}_{d^5}(\ell) = f_0 + \frac{f_1}{\ell^{5/2}} \,.
\ee
With this last fit we find
\be
\boxed{{\cal C}_{d^5} = {\cal C}_{d^5}(\infty) = 0.3060} \,.
\label{d5Feynman} \ee

\subsection{The contact term}
\label{s_d5contact}

The multilinear $N$-string function at genus zero is given by \cite{Zwie-BV, Belo-Zwie}
\be
\left\{ \Psi_1,\ldots,\Psi_N\right\} = \left(\frac{i}{2 \pi}\right)^{N-3} 
\int_{{\cal V}_{0,N}} d\lambda_1 \wedge \ldots \wedge d\lambda_{2(N-3)} \langle \Sigma| 
b(v_{\lambda_1}) \ldots b(v_{\lambda_{2(N-3)}}) |\Psi_1\rangle \ldots 
|\Psi_N\rangle \,.
\ee
The $\lambda_i$ are $2(N-3)$ real coordinates of the reduced moduli space ${\cal V}_{0,N}$ 
of the $N$-punctured spheres, and the surface state $\langle \Sigma|$ corresponds 
to the $N$-punctured sphere given by the parameters $\lambda_i$.
The antighost insertions $b(v_{\lambda_i})$ are given by
\be
b(v_{\lambda_i}) = \sum_{I=1}^N \sum_{m=-1}^\infty \left( H_{i,m}^I b_m^{(I)} + 
\overline{H_{i,m}^I} \bar{b}_m^{(I)} \right) \ , \quad {\rm where} \quad 
H_{i,m}^I = \oint \frac{dw}{2 \pi i} \frac{1}{w^{m+2}} \frac{1}{h_I'} 
\frac{\partial h_I}{\partial \lambda_i} \,.
\label{binsertions} \ee
The $h_I(w_I; \lambda_1, \ldots \lambda_{2(N-3)})$ are the $N$ maps from the local 
coordinates $w_I$ to the sphere, which are going to be described a little later. 
Now we follow \cite{Yang:2005ep} and rewrite the antighost insertions in a more 
convenient form. First we rename the coordinates of the moduli space as 
\be
\xi_i = x_i + i \, y_i \ , \quad {\rm where} \quad 
x_i = \lambda_{2i-1} \ , \ y_i = \lambda_{2i} \ , \ i=1,\ldots,N-3 \,,
\ee
where the $\xi_i$ are the complex coordinates naturally used in the quadratic differentials 
formalism. We have
\be
dx_i \wedge d y_i \, b(v_{x_i}) b(v_{y_i}) = 
d\xi \wedge d\bar{\xi} \, {\cal B}_i {\cal B}_i^\star 
= -2 i \, dx_i \wedge d y_i \, {\cal B}_i {\cal B}_i^\star \,,
\ee
where
\be
{\cal B}_i = \sum_{I=1}^N \sum_{m=-1}^\infty \left(B_{i,m}^I b_m^{(I)} + 
\overline{C_{i,m}^I} \bar{b}_m^{(I)} \right) \quad , \quad 
{\cal B}_i^\star = \sum_{I=1}^N \sum_{m=-1}^\infty \left(C_{i,m}^I b_m^{(I)} + 
\overline{B_{i,m}^I} \bar{b}_m^{(I)} \right) \,,
\ee
and the coefficients $B_{i,m}^I$ and $C_{i,m}^I$ are given in terms of derivatives 
with respect to $\xi_i$ and $\bar{\xi_i}$
\be
B_{i,m}^I = \oint \frac{dw}{2 \pi i} \frac{1}{w^{m+2}} \frac{1}{h_I'} 
\frac{\partial h_I}{\partial \xi_i} \quad , \quad 
C_{i,m}^I = \oint \frac{dw}{2 \pi i} \frac{1}{w^{m+2}} \frac{1}{h_I'} 
\frac{\partial h_I}{\partial \bar{\xi_i}} \,.
\label {BC} \ee
And we arrive at the formula
\be
\left\{ \Psi_1,\ldots,\Psi_N\right\} = \frac{1}{\pi^{N-3}} 
\int_{{\cal V}_{0,N}} dx_1 \wedge dy_1 \wedge \ldots \wedge dx_{N-3} \wedge dy_{N-3} 
\langle \Sigma| \left({\cal B} {\cal B}^\star \right)_1 \ldots 
\left({\cal B} {\cal B}^\star \right)_{N-3} |\Psi_1\rangle \ldots |\Psi_N\rangle \,.
\label{N-amplitude} \ee
We will now consider the case of interest, $N=5$ and $|\Psi_i \rangle = |D\rangle$, 
where $|D\rangle$ is the dilaton (\ref{dilaton}).
we will thus only need the coefficients $B_{i,m}^I$ and $C_{i,m}^I$ with $m=-1$ and $m=1$. 
To evaluate them, we need to expand the maps $h_I$ to order $w_I^3$
\be
z = h_I(w_I;\xi_1, \bar{\xi_1}, \xi_2, \bar{\xi_2}) = z_I + 
\rho_I \, w_I + \rho_I^2 \beta_I \, w_I^2 + \rho_I^3 \gamma_I \, w_I^3 + {\cal O}(w_I^4) \,,
\label{h-expand}
\ee
where all the coefficients on the right-hand side depend on $\xi_1$, $\bar{\xi_1}$, $\xi_2$ 
and $\bar{\xi_2}$. The $z_I$ are the positions of the finite poles on the $z$-plane, namely
\be
z_1 = 0 \quad , \quad z_2 = 1 \quad , \quad z_3 = \xi_1 \quad , \quad z_4 = \xi_2 \,.
\ee
For the puncture at infinity, we must use the coordinate $t = 1/z$
\be
t = h_5(w_5;\xi_1, \bar{\xi_1}, \xi_2, \bar{\xi_2}) = 
\rho_5 \, w_5 + \rho_5^2 \beta_5 \, w_5^2 + \rho_5^3 \gamma_5 \, w_5^3 + {\cal O}(w_5^4) \,.
\ee
We can now use (\ref{BC}) to express the coefficients that we need. We find
\ben
&& \hspace{-24pt} B_{1,-1}^I = \frac{1}{\rho_3} \delta_{3,I} \quad , \quad 
B_{2,-1}^I = \frac{1}{\rho_4} \delta_{4,I} \quad , \quad 
C_{i,-1}^I = 0 \nonumber \\
&& \hspace{-24pt} B_{1,1}^I = \rho_I \frac{\partial \beta_I}{\partial \xi_1} + 
\frac{1}{2} \rho_3 \epsilon_3 \delta_{I,3} \ , \quad 
B_{2,1}^I = \rho_I \frac{\partial \beta_I}{\partial \xi_2} + 
\frac{1}{2} \rho_4 \epsilon_4 \delta_{I,4} \ , \quad
C_{i,1}^I = \rho_I \frac{\partial \beta_I}{\partial \bar{\xi_i}} \,,
\label{BBCC} \een
where
\be
\epsilon_I \equiv 8 \beta_I^2-6 \gamma_I \,. 
\ee
The coefficients $\beta_I$ and $\gamma_I$ can be calculated from the expression of the 
quadratic differential (\ref{phi5}) and (\ref{h5_2}). First we expand 
$\varphi = \phi(z) (dz)^2$ around the puncture $z_I$
\be
\phi(z) = -\frac{1}{(z-z_I)^2} + \frac{r_{-1}^I}{(z-z_I)} + r_0^I + {\cal O}(z) \,,
\ee
and knowing that in the local coordinates $\varphi$ takes the canonical form 
\be
\varphi = - \frac{1}{w_I^2} (dw_I)^2 \,,
\ee
we find the relations
\be
\beta_I = \frac{1}{2} r_{-1}^I \quad , \quad \gamma_I = \frac{1}{16} \left(
7 (r_{-1}^I)^2 + 4 (r_0^I)^2 \right) \,.
\ee
And after expanding $\phi(z)$ to obtain $r_{-1}^I$ and $r_0^I$, we find
\ben
\beta_1 &=& \frac{a_1 - \xi_1^2 - \xi_2^2}{2 \xi_1 \xi_2} \nonumber \\
\beta_2 &=& -\frac{a_1+a_2+2 \xi_1 - \xi_1^2 + 2 \xi_2 - \xi_2^2}
{2 (\xi_1-1) (\xi_2-1)} \nonumber \\
\beta_3 &=& \frac{a_1-2 \xi_1^2 - \xi_2^2 + \xi_1 (3+a_2+2\xi_2)}
{2 \xi_1 (\xi_1-1) (\xi_2-\xi_1)} \nonumber \\
\beta_4 &=& \frac{a_1 - \xi_1^2 + 2 \xi_1 \xi_2 + \xi_2(3+a_2-2 \xi_2)}
{2 \xi_2 (\xi_2-1) (\xi_1-\xi_2)} \nonumber \\
\beta_5 &=& -\frac{1}{2} \left(2 + a_2 + \xi_1 + \xi_2 \right)
\label{beta} \,, \een
and for $\epsilon_I = 8\beta_I^2 - 6 \gamma_I$, we find
\ben
\epsilon_1 \hspace{-4pt} &=& \hspace{-4pt} \frac{-5(u-a_1)^2+12\left((\xi_1^3+\xi_2^2) 
(1+\xi_2)-a_1(s+t)+\xi_1^2(1-\xi_2+\xi_2^2) + 
t(-3-a_2-\xi_2+\xi_2^2)\right)}{8 t^2} \nonumber \\
\epsilon_2 \hspace{-4pt} &=& \hspace{-4pt} \frac{-5(u-2s-a_1-a_2)^2 - 12\left(5s-7u+2w+t(4s-5-u-t) - 
a_2(s-2)-a_1(2s-t-3)\right)}{8(t-s+1)^2} \nonumber \\
\epsilon_3 \hspace{-4pt} &=& \hspace{-4pt} \frac{1}{8(\xi_1(\xi_1-\xi_2)(\xi_1-1))^2} 
\left(4\xi_1^3(10\xi_1-15-a_2-14\xi_2) - 
a_1(5a_1+16\xi_1^2+2\xi_1(3+5a_2-2\xi_2) \right.
\nonumber \\
&& \hspace{4cm} + 2(6-5\xi_2)\xi_2) + 
\xi_2^2(12(1+\xi_2)-5\xi_2^2)-2t(12+9\xi_2-5a_2\xi_2+2\xi_2^2) \nonumber \\ 
&& \hspace{4cm} \left. + \xi_1^2(15-18a_2-5a_2^2+24\xi_2-8a_2\xi_2+44\xi_2^2) \right) 
\nonumber \\
\epsilon_4 \hspace{-4pt} &=& \hspace{-4pt} \frac{-1}{8 (\xi_2 (\xi_1-\xi_2) (\xi_2-1))^2} \left(
- a_1(-5 a_1+10\xi_1^2-12\xi_1+4t+2(-5a_2-3-8\xi_2)\xi_2) 
\right. \nonumber \\
&& \hspace{4cm} + 2\xi_1^2(-6+(9-5a_2)\xi_2-22\xi_2^2) + 8t(3-(3-a_2)\xi_2+7\xi_2^2) 
\nonumber \\
&& \hspace{4cm} \left. - \xi_2^2(15-18a_2-5a_2^2-60\xi_2-4a_2\xi_2+40\xi_2^2) + 
\xi_1^3(5 \xi_1-12+4\xi_2)\right) \nonumber \\
\epsilon_5 \hspace{-4pt} &=& \hspace{-4pt} \frac{1}{8} 
\left(16+12a_1+4s+7u+2t+a_2(-5a_2-8+2s) \right) \,,
\label{epsilon} \een
where we have defined 
\be
s \equiv \xi_1+\xi_2 \ , \quad t \equiv \xi_1 \xi_2 \ , \quad 
u \equiv \xi_1^2+\xi_2^2 \ , \quad w \equiv \xi_1^3+\xi_2^3 \,.
\ee
We will also need to compute a few correlators. To see which ones, it is best 
to start expanding the antighost insertions. It is easily seen from the expression 
for the dilaton and the fact that only the correlators with ghost number $(3,\bar{3})$ 
do not vanish, that we need either three $b$'s and one $\bar{b}$, or one $b$ and 
three $\bar{b}$'s; and as already mentioned we only need antighost oscillators with 
indices $-1$ or $1$. Let us expand $\left({\cal B}{\cal B}^{\star}\right)_1 
\left({\cal B}{\cal B}^{\star}\right)_2$ in the following way: we take all terms in the 
first factor that have either two $b$'s or two $\bar{b}$'s and mixed terms in the 
second factor. The rest of the expression, with mixed terms in the first factor, can be 
found simply by changing some indices 
\ben
&& \left({\cal B}{\cal B}^{\star}\right)_1 \left({\cal B}{\cal B}^{\star}\right)_2 =
\nonumber \\
&=& \left( B_{1,-1}^3 b_{-1}^{(3)} \sum_{I=1}^5 C_{1,1}^I b_1^{(I)} - 
\overline{B_{1,-1}^3} \bar{b}_{-1}^{(3)} \sum_{I=1}^5 \overline{C_{1,1}^I} 
\bar{b}_1^{(I)} + \sum_{I\neq J} B_{1,1}^I C_{1,1}^J b_1^{(I)} b_1^{(J)} - 
\sum_{I\neq J} \overline{B_{1,1}^I} \overline{C_{1,1}^J} \bar{b}_1^{(I)} 
\bar{b}_1^{(J)} \right) \nonumber \\
&& \times \left( B_{2,-1}^4 b_{-1}^{(4)} \sum_{I \neq 4} \overline{B_{2,1}^I} 
\bar{b}_1^{(I)} - \overline{B_{2,-1}^4} \bar{b}_{-1}^{(4)} \sum_{I\neq 4} 
B_{2,1}^I b_1^{(I)} + 
\sum_{I \neq J} M_2^{IJ} b_1^{(I)} \bar{b}_1^{(J)} \right) + 
\left( {}_1 \leftrightarrow {}_2 \ , \ {}^3 \leftrightarrow {}^4 \right) \,, 
\nonumber \\
\label{BBBB} \een
where 
\be
M_i^{IJ} \equiv B_{i,1}^I \overline{B_{i,1}^J} - C_{i,1}^I \overline{C_{i,1}^J} \,,
\ee
and the last term is obtained by changing as indicated the left subscripts 
of $B$ and $C$, the subscript  of $M$, and all superscripts. Noting that 
\be
b_{-1}|D\rangle = c_{-1}|0\rangle \ , \quad 
b_1|D\rangle = -c_1|0\rangle \ , \quad
\bar{b}_{-1}|D\rangle = -\bar{c}_{-1}|0\rangle \ , \quad
\bar{b}_1|D\rangle = \bar{c}_1|0\rangle \,,
\ee
we see from (\ref{BBBB}), that we need the following open correlators
\ben
&& A_{IJ} \equiv \langle (c_{-1} c_1)^{(I)}, c_{-1}^{(J)} \rangle_o \ , \quad 
B_{IJ} \equiv \langle (c_{-1} c_1)^{(I)}, c_1^{(J)} \rangle_o \label{correlators} \\
&& C_{IJK} \equiv \langle c_1^{(I)}, c_1^{(J)}, c_1^{(K)} \rangle_o \ , \quad 
D_{IJK} \equiv \langle c_{-1}^{(I)}, c_1^{(J)}, c_1^{(K)} \rangle_o \ , \quad 
E_{IJK} \equiv \langle c_{-1}^{(I)}, c_{-1}^{(J)}, c_1^{(K)} \rangle_o \,.
\nonumber
\een
The conventions for the closed correlators are as in \cite{Yang:2005ep}
\be
\langle c(z_1) c(z_2) c(z_3) \bar{c}(\bar{w}_1) \bar{c}(\bar{w}_2) 
\bar{c}(\bar{w}_3) \rangle
= -2 \langle c(z_1) c(z_2) c(z_3) \rangle_o \, \langle \bar{c}(\bar{w}_1) 
\bar{c}(\bar{w}_2) \bar{c}(\bar{w}_3) \rangle_o \,,
\ee
and the open correlator is 
\be
\langle c(z_1) c(z_2) c(z_3) \rangle_o = (z_1-z_2) \, (z_1-z_3) \, (z_2-z_3) \,.
\ee
$A_{IJ}$ and $B_{IJ}$ were already calculated in
\cite{Yang:2005ep}. For the other correlators, we can use either the
conformal transformation of the operator $\frac{1}{2} \partial^2c(z)$
corresponding to $c_{-1}$ (as in \cite{Yang:2005ep}), or conservation
laws as in \cite{Moe-Yang}. We find when $I,J,K \neq 5$ (with the
definition $Z_{IJ} \equiv z_I-z_J$):
\ben
A_{IJ} &=& \rho_J \left( \beta_J - \beta_I - 2 \beta_I \beta_J z_{IJ} 
+ \frac{1}{2} \epsilon_J z_{IJ} (1-\beta_I z_{IJ}) \right) \nonumber \\
B_{IJ} &=& \frac{1}{\rho_J} z_{IJ} (1-\beta_I z_{IJ}) \nonumber \\
C_{IJK} &=& \frac{1}{\rho_I \rho_J \rho_K} z_{IJ} z_{IK} z_{JK} \label{corrnot5} \\
D_{IJK} &=& \frac{\rho_I}{\rho_J \rho_K} \left( z_{JK} - 
\beta_I (z_{IK} + z_{IJ}) z_{JK} + 
\frac{1}{2} \epsilon_I z_{IJ} z_{JK} z_{IK} \right) \nonumber \\
E_{IJK} &=& \frac{\rho_I \rho_J}{\rho_K} \left( \beta_I - \beta_J + 
\beta_I \beta_J (z_{IJ} + z_{IK} - z_{JK}) + \frac{1}{2} \epsilon_J z_{JK} - 
\frac{1}{2} \beta_I \epsilon_J (z_{IJ} + z_{IK}) z_{JK} + \right. \nonumber \\
&& \hspace{1.2cm} \left. + \frac{1}{2} \epsilon_I \left(-z_{IK} + \beta_J z_{IK} (z_{JK} - z_{IJ}) 
+ \frac{1}{2} \epsilon_J z_{IJ} z_{IK} z_{JK} \right) \right)
\quad , \quad I,J,K \neq 5 \,. \nonumber
\een
The cases when oscillators are at infinity must be treated
separately. Since $B_{i,-1}^5$ and $C_{i,-1}^5$ are zero, we cannot
have a $c_{-1}$ alone at infinity. We therefore need
\ben
A_{5J} &=& \rho_J \left( \frac{1}{2} \epsilon_J (\beta_5 + z_J) - \beta_J \right) \nonumber \\
B_{I5} &=& \frac{\beta_I}{\rho_5} \quad , \quad B_{5J} = \frac{1}{\rho_J} (z_J + \beta_5) 
\nonumber \\
C_{IJ5} &=& \frac{z_{JI}}{\rho_I \rho_J \rho_5} \nonumber \\
D_{IJ5} &=& \frac{\rho_I}{\rho_J \rho_5} \left( \beta_I - \frac{1}{2} \epsilon_I z_{IJ}
\right) \nonumber \\
E_{IJ5} &=& \frac{\rho_I \rho_J}{\rho_5} \left( \frac{1}{2} \beta_I \epsilon_J - 
\frac{1}{2} \epsilon_I \beta_J - \frac{1}{4} \epsilon_I \epsilon_J z_{IJ} \right) \,,
\label{corr5} \een
and we note that $C_{IJK}$ is totally antisymmetric, $D_{IJK}$ is antisymmetric in 
$J$ and $K$, and $E_{IJK}$ is antisymmetric in $I$ and $J$.

\paragraph{}
We can now write the integrand of (\ref{N-amplitude}) with five dilatons from 
(\ref{BBBB}) and the definitions (\ref{correlators}).
\ben
&& \langle \Sigma | \left({\cal B}{\cal B}^{\star}\right)_1 
\left({\cal B}{\cal B}^{\star}\right)_2 
|D\rangle |D\rangle |D\rangle |D\rangle |D\rangle = 
\nonumber \\
&=& 4 \Re \left\{ \frac{1}{\rho_3 \rho_4} \sum_{3\neq 4\neq I\neq J \neq K} \Bigl\{ 
\overline{A_{k3}} \left( 
D_{4IJ} C_{2,1}^I B_{1,1}^J - B_{IJ} C_{2,1}^4 B_{1,1}^J 
+ B_{JI} C_{2,1}^I B_{1,1}^4 \right) \right.
\begin{array} {c} \\ \\ \end{array}
\nonumber \\
&& \hspace{1cm} + \overline{A_{K4}} \left( D_{3IJ} C_{1,1}^I B_{2,1}^J - B_{IJ} C_{1,1}^3 B_{2,1}^J 
+ B_{JI} C_{1,1}^I B_{2,1}^3 \right) 
\nonumber \\
&& \left. \hspace{-1cm} + \overline{B_{KJ}} 
\left( -E_{34I} \left( C_{1,1}^I \overline{B_{2,1}^J} - C_{2,1}^I \overline{B_{1,1}^J} \right) 
+ \overline{B_{2,1}^J} \left( A_{I3} C_{1,1}^4 - A_{I4} C_{1,1}^3 \right) + 
\overline{B_{1,1}^J} \left( A_{I4} C_{2,1}^3 - A_{I3} C_{2,1}^4 \right) \right) \right\}
\nonumber \\
&& + \frac{1}{\rho_3} \sum_{3 \neq I \neq J \neq K \neq L} 
\Bigl\{ \overline{B_{LK}} \Bigl( -D_{3IJ} C_{1,1}^I M_2^{JK} 
+ B_{IJ} C_{1,1}^3 M_2^{JK} - B_{JI} C_{1,1}^I M_2^{3K}
\nonumber \\
&& \hspace{1cm} \left. \left. + \overline{B_{1,1}^K} \left( 
-D_{3IJ} B_{2,1}^I C_{2,1}^J + B_{IJ} B_{2,1}^3 C_{2,1}^J - 
B_{JI} B_{2,1}^I C_{2,1}^3 \right) \right) - C_{IJK} 
\overline{A_{L3}} B_{2,1}^I C_{2,1}^J B_{1,1}^K \right\}
\nonumber \\
&& + \frac{1}{\rho_4} \sum_{4 \neq I \neq J \neq K \neq L} 
\Bigl\{ \overline{B_{LK}} \left( -D_{4IJ} C_{2,1}^I M_1^{JK} 
+ B_{IJ} C_{2,1}^4 M_1^{JK} - B_{JI} C_{2,1}^I M_1^{4K} \right.
\nonumber \\
&& \hspace{1cm} \left. \left. + \overline{B_{2,1}^K} \left( 
-D_{4IJ} B_{1,1}^I C_{1,1}^J + B_{IJ} B_{1,1}^4 C_{1,1}^J - 
B_{JI} B_{1,1}^I C_{1,1}^4 \right) \right) - C_{IJK} 
\overline{A_{L4}} B_{1,1}^I C_{1,1}^J B_{2,1}^K \right\}
\nonumber \\
&& \left. + \sum_{I \neq J \neq K \neq L \neq T} C_{IJK} 
\overline{B_{TL}} \left(B_{1,1}^I C_{1,1}^J M_2^{KL} + 
B_{2,1}^I C_{2,1}^J M_1^{KL} \right) \right\} \,.
\een
With (\ref{BBCC}), (\ref{beta}), (\ref{epsilon}), (\ref{corrnot5}) and 
(\ref{corr5}), this expression can be expressed in terms of $\xi_i$, 
$a_i$, $\partial a_i / \partial \xi_j$, 
$\partial a_i / \partial \bar{\xi_j}$ and $\rho_I$, and we can therefore 
integrate it numerically over the reduced moduli space of five-punctured 
spheres. Since we have five times the same state, the result is simply 
120 times the integral on ${\cal A}_5$, thus
\be
\kappa^2 V_{d^5} = \frac{1}{5!} \left\{D,D,D,D,D\right\} = 
\frac{1}{\pi^2} \int_{{\cal A}_5} 
dx_1 \wedge dy_1 \wedge dx_2 \wedge dy_2 \langle \Sigma| \left( {\cal B} 
{\cal B}^\star \right)_1 \left( {\cal B} {\cal B}^\star \right)_2 
|D\rangle |D\rangle |D\rangle |D\rangle |D\rangle \,.
\label{d5-int} \ee
We do the integration as in Section \ref{s_t5}, on three different grids. The results 
are shown in Table \ref{d5-table}.
\begin{table}[!ht]
\begin{center}
\begin{tabular}{|c||c|c|c|}
\hline
$(N,h,s)$ & $(30,0.1,0.1)$ & $(70,0.05,0.05)$ & $(70,0.05,0.04)$ \\
\hline \hline
$\kappa^2 V_{d^5}(2h)$ & $-0.30666  \pm 0.0093$ & $-0.30768 \pm 0.0023$ & $-0.30759 \pm 0.0016$ \\
\hline
$\kappa^2 V_{d^5}(h)$ & $-0.30343 \pm 0.0093$ & $-0.30666 \pm 0.0023$ & $-0.30660 \pm 0.0016$ \\
\hline \hline
$\kappa^2 V_{d^5}$ & $-0.3024 \pm 0.0093$ & $-0.3063 \pm 0.0023$ & $-0.3063 \pm 0.0016$ \\
\hline
\end{tabular}
\caption{\footnotesize{The results of the integration (\ref{d5-int})
and its uncertainty on three different grids.}}
\label{d5-table}
\end{center}
\end{table}
They are compatible within their error bounds, and we take again the result from 
the finest grid as final answer
\be
\boxed{\kappa^2 V_{d^5} = -0.3063 \pm 0.0016} \,.
\label{d5contact} \ee
We see that (\ref{d5Feynman}) and (\ref{d5contact}) cancel each other
with a precision of about $0.1\%$, well within the error bound of
$0.5\%$ on the contact term (\ref{d5contact}). This is solid
evidence that our computations are reliable.

\sectiono{Conclusions and prospects}
\label{s_conclusions}

In the light of the verification successfully made in Section
\ref{s_d5}, that the effective potential of the dilaton vanishes at
order five, we can claim that the techniques described in this paper
work well, that the reduced moduli space is understood and described
right, and that our implementation of the algorithm gives reliable
results. In particular we trust the value (\ref{t5}) obtained for the
contact term of five tachyons. It is also a good check of the
consistency of CSFT itself.

\paragraph{}
We are able to estimate the uncertainty made in the computation of
terms (see (\ref{t5}) and (\ref{d5contact})). However we want to
mention here that we have been quite conservative in this
estimation. In particular the errors on the fits were treated as
systematic instead of independent. This is only half right because we
expect interpolations made from different samples to be more or less
independent, although two interpolations at two nearby points using
the same sample are not independent. The overestimation of the error
seems confirmed by the values in Tables \ref{t5-table} and
\ref{d5-table}. However one should be careful in this respect because,
as we already mentioned, we don't think that the errors made on
different grids are independent either. So unless we find a better way
of estimating the error, we will stick to our conservative
estimation. Any way, we emphasize that the errors on any given term
will probably be no more than $0.5\%$, the error we find for the
five-dilaton term (which is calculated from a very long
expression). And this precision is probably enough to do level
truncation.

\paragraph{}
The effect of the five-tachyon term (\ref{t5}) on the stable vacuum is
studied in \cite{Moe-Yang}. However this term of level zero isn't
enough to draw any conclusion. For this, it will be necessary to
compute other terms at higher level. At level two there is only one
term, namely four tachyons and one dilaton $V_{t^4d}$. At level four
we'll have five terms, $V_{t^3d^2}$ and $V_{t^4\psi_i}$ where
$|\Psi_i\rangle$, $i=3,\ldots,6$, are the four scalar fields at level
four. The main difficulty in computing these terms will be that when
we have different external states, the integrations on the $120$
pieces of reduced moduli space won't be all equal, and it will require
some (straightforward but lengthy) work to express them as one
integral over ${\cal A}_5$; the extreme case being when we have five
different states, we will have $60$ different integrals (complex
conjugation is always trivial and divides the number of integrations
by two). But the computation of interactions to level four is
certainly not that bad. By automatizing the computation of oscillator
algebra and correlators, as in \cite{Moe-Yang}, it might be possible
to compute higher levels as well.

\paragraph{}
At last, it would be useful to make our numerical data available, so
that readers can use it to make their own computations of quintic
terms. Ideally we would like to create fits of the boundaries of the
moduli space and the parameters of the quadratic differentials $a_1$
and $a_2$ and the mapping radii, that could hold in a paper and be
entered in a computer in a reasonable amount of time. This is under
study, but we are a bit pessimistic given the dimensionality and
rather complicated shape of the reduced moduli space (our data
describing its boundary is about 600 megabytes large).

\section*{Acknowledgments}
I am indebted to Barton Zwiebach for many useful discussions. This
work has been funded by an "EC" fellowship within the framework of the
"Marie Curie Research Training Network" Programme, Contract
no. MRTN-CT-2004-503369.


\end{document}